\documentclass[a4paper,11pt]{article}
\pdfoutput=1 

\usepackage{jcappub} 

\usepackage{caption}
\usepackage{soul}
\usepackage{amsmath}
\usepackage{amsmath}
\usepackage{rotating}
\usepackage{tabularx}
\usepackage{graphicx}
\usepackage{color}
\usepackage{natbib}
\usepackage{longtable,lscape}
\usepackage{txfonts}
\usepackage{amsfonts}
\usepackage{amssymb}
\usepackage{newlfont}
\usepackage{textcomp}
\usepackage{times}
\usepackage{units}
\usepackage{mathbbol}
\usepackage[switch]{lineno}
\usepackage{epstopdf}
\usepackage{ifthen}

\def\ba{\begin{eqnarray}}
\def\ea{\end{eqnarray}}
\def\be{\begin{equation}}
\def\ee{\end{equation}}
\newcommand{\comm}[1]{}
\renewcommand{\comm}[1]{ \textcolor{magenta}{#1}}

\newcommand{\ms}{\scriptscriptstyle}
\title{\boldmath Extended reionization in models beyond $\Lambda$CDM with Planck 2018 data}

\author[1,2]{Daniela Paoletti,}
\author[3,1,2]{Dhiraj Kumar Hazra,}
\author[1,2]{Fabio Finelli,}
\author[4,5,6,7]{George F. Smoot}
\affiliation[1]{Osservatorio di Astrofisica e Scienza dello Spazio di Bologna/Istituto Nazionale di Astrofisica, via Gobetti 101, I-40129 Bologna, Italy}
\affiliation[2]{Istituto Nazionale Di Fisica Nucleare, Sezione di Bologna,Viale Berti Pichat, 6/2, I-40127 Bologna, Italy}
\affiliation[3]{The Institute of Mathematical Sciences, HBNI, CIT Campus, Chennai 600113, India}
\affiliation[4]{Paris Centre for Cosmological Physics, Universit\'e de Paris,
CNRS, Astroparticule et Cosmologie, F-75013 Paris, France}
\affiliation[5]{Institute for Advanced Study \& Physics Department, Hong Kong University of Science and Technology, Clear Water Bay, Kowloon,Hong Kong}
\affiliation[6]{Physics Department and Lawrence Berkeley National Laboratory, University of California, Berkeley, CA 94720, USA}
\affiliation[7]{Energetic Cosmos Laboratory, Nazarbayev University, Astana, Kazakhstan}
\emailAdd{daniela.paoletti@inaf.it, dhiraj@imsc.res.in, fabio.finelli@inaf.it, gfsmoot@lbl.gov} 

\abstract{We provide an update on the constraints on extended reionization histories
with the Planck 2018 cosmic microwave background anisotropy data.
The Planck 2018 data on large angular scales improve the measurement of
the $E$-mode polarization reionization bump at low multipoles providing the
possibility to improve our previous results~\cite{HPFS18}.
Using a minor modification to the original Poly-reion model~\citep{HS17}
for the reionization history, we find that the Planck 2018 data significantly improve
all our previous results: we find as optical depth of $\tau=0.0572_{\ms -0.0075}^{\ms{+0.0064}}$
at 68\% CL, that early onsets of reionization are strongly disfavoured, 
i.e. redshift when the reionization begins, $z_{xe=0}=18.18_{\ms-10.89}^{\ms +1.61}$ at 68\% CL,
and that reionization duration (defined between 10\% and 99\% reionization) is significantly reduced, i.e. $\Delta_z^{Reion}=4.59_{\ms-2.45}^{\ms+1.67}$ at 68\% CL.
We explore possible correlations between reionization histories and cosmological parameters, including important extensions beyond $\Lambda$CDM. We find that the degeneracy
between reionization and scalar spectral index,
neutrino mass sum, spatial curvature, dark matter annihilation
and other non-standard models are significantly reduced.The reduction of the error bars and the degeneracies, together with the shift towards lower values of the optical depth that we observe in the Poly-reion model are mainly driven by the new low-$\ell$ polarization likelihood of Planck 2018 baseline based on the HFI data. This is confirmed also by the results derived without this likelihood and the ones with different alternatives to the baseline that are presented for a subset of models.
}

\begin{document}
\maketitle
\section{Introduction}
Reionization marks the end of the dark ages, with the first sources of ionizing photons lighting up.
The CMB anisotropy pattern is a main source of information on the reionization process.
Its sensitivity to the details of reionization originates from two different effects: 1) the opacity
generated by reionization suppresses
the power of CMB anisotropies because of the rescattering of photons (this effect is at the basis of the
degeneracy between the amplitude of primordial fluctuations and the optical depth); 2) the rescattering of
the photon quadrupole generates a bump at the multipoles below ten in the CMB polarization anisotropies.
The reionization bump in the E-mode polarization provides information on both the value of the (average)
optical depth, that is related to its amplitude, and to a lesser extent on
the duration of reionization, that is instead
related to the shape of the bump.

The latest results in CMB polarization anisotropy measurements by Planck \cite{Aghanim:2016yuo,Adam:2016hgk,Aghanim:2018eyx} have shed light on
the reionization process. The Planck 2018 determination of the (average) optical depth is \cite{Aghanim:2018eyx}
$\tau = 0.054 \pm 0.007$ (TTTEEE+lowE+lensing) at 68 \% CL.
Being the average optical depth the one of the six cosmological parameter of the $\Lambda$CDM cosmology with the
largest uncertainty, it is a standard approximation to compress the CMB information on reionization
in the (effective) redshift at which it occurred by adopting an hyperbolic tangent variation in the ionization fraction, which is therefore determined as $z_\mathrm{re}$ at 68 \%CL.

With these recent CMB polarization measurements the constraints on reionization beyond the simple hyperbolic
tangent model used as baseline in cosmology have also significantly improved.
These investigations include physical models for the reionization history \cite{Adam:2016hgk}
or a pure free form data fitting with principal component analysis~\cite{Heinrich:2016ojb,Heinrich:2018btc}, to hybrid
parametric fitting models~\cite{HS17,Millea:2018bko,Aghanim:2018eyx} and astrophysical data driven
reconstructions~\cite{Hazra:2019wdn,Gorce,REIO}.

The analysis of CMB anisotropy data beyond the optical depth are not relevant only for the understanding  
of the physics of reionization, 
because of its known and possible degeneracy with other cosmological parameters \cite{Efstathiou:1998xx,Howlett:2012mh}, as
for instance with the amplitude of the power spectrum of primordial perturbations.
The situation gets more complex when we open up to extensions of the standard cosmological
model as in~\cite{HPFS18}, where the degeneracy with neutrino sector and DM annihilation
are studied. The same situation arises also for the study of alternative choices of the primordial
power spectrum as in the search for primordial features, where as shown in~\citep{HPBFSSS17,Mortonson:2009qv}
the uncertainty on the reionization process diminishes the significance of possible detections.

In this work we update with Planck 2018 data our previous constraints on the parametric model of
reionization {\it termed} Poly-reion~\cite{HPFS18} not only within $\Lambda$CDM, but also within
several extended cosmological models to search for potential degeneracies
of the physics of reionization with extra cosmological parameters beyond $\Lambda$CDM.
In addition to Planck 2018 data, we also use the reprocessing of large scale Planck HFI polarization data from~\cite{Delouis:2019bub}.

The article is organized as follows. In section 2 we describe the Poly-reion model in the form
used in this work. In section 3 we present the results for the standard cosmological model with
different combination of Planck 2018 likelihoods and
the more recent reprocessing of Planck HFI large scale polarization data from Planck~\cite{Delouis:2019bub}.
In section 4 we present the results of the Planck 2018
baseline for the main extensions to the standard cosmological model. We conclude in Section 5.

\flushbottom 
\section{Poly-reion model for the history of reionization}

In order to go beyond the standard hyperbolic tangent model for $x_e$,we use the Poly-reion model of reionization. Poly-reion was originally introduced in~\cite{HS17}, it interpolates the ionization fraction between different nodes; in its original version it assumed fixed nodes at different redshifts, thereafter instead it used flexible nodes as in~\cite{HPBFSSS17,HPFS18}. Poly-reion uses Piecewise Cubic Hermite Interpolating Polynomial (PCHIP)~\cite{PCHIP} to interpolate the free electron fractions ($x_e(z)$) between different nodes in the redshift space. We use four nodes, the first two are fixed by assuming that the reionization is completed by redshift ($z$) 5.5~\footnote{While in most cases $z=6$ is considered to be the completion of reionization, we use 5.5 to be conservative following analyses that suggest the possible presence of reionization at $z<6$~\cite{Schroeder:2012uy}} and the Universe remained completely ionized till date~\cite{McGreer:2014qwa}, namely we fix the $z=0$ and $z=5.5$ nodes. The other two nodes consider a flexible intermediate node called $z_{\mathrm int}$ and $z_{xe=0}$ which denotes the redshift where the reionization begins and $x_e=0$. In the intermediate node we vary both the position in redshift and the amplitude of the ionization fraction: $z_{\mathrm int}$ is allowed to vary between 5.5 and 30 and $x_e(z)$ takes the values between 0 and 1. Given $z_{\mathrm int}$ and $x_e(z_{\mathrm int})$, we solve for $z_{xe=0}$ for a value of optical depth $\tau$ between $z_{\mathrm int}$ and a very high redshift ($z=70$ in our case). Together with the position and amplitude of the intermediate node $z_{\mathrm int}$ and $x_e(z_{\mathrm int})$,  the optical depth represents the third parameter describing the reionization in our model and we vary it with a uniform prior. 

By construction this model of Poly-reion allows only monotonic reionization, in fact, the use of PCHIP in the interpolation guarantees the consistency of the electron fraction which can vary only between the physical values 0 and 1 as this Polynomial preserves the form. We note anyway that monotonic histories have been shown to be the most favoured by CMB and astrophysical data~\cite{Hazra:2019wdn}.
The reionization history constructed with Poly-reion is used only for hydrogen reionization in our work.
We assume that Hydrogen reionization is completed by $z=5.5$ concerning the other two steps of reionization regarding Helium I and Helium II, we assume the standard hyperbolic tangent form with the first helium reionization at the same time as the hydrogen and the second at redshift of 3.5 as in Planck baseline. 
The standard model assumed for reionization is the hyperbolic tangent whose standard form assumes:
\begin{equation}
    x_e(z)=\frac{1+F_{\mathrm He}}{2}\left[1+\tanh\left(\frac{y(z_{\mathrm re})-y(z)}{\delta_{\mathrm reion}}\right)\right],
\end{equation}
where $y(z)=(1+z)^{3/2}$, $\delta_{\mathrm reion}=1.5\sqrt{1+z_{\mathrm re}}dz$. $z_{\mathrm re}$ denotes the redshift at which this symmetric model of reionization is halfway complete. $F_{\mathrm He}$ is calculated from mass fraction of helium consistent with Big Bang Nucleosynthesis. Standard analysis assumes $dz=0.5$ which is also used in the $Planck$ grid results.\footnote{\url{http://www.cosmos.esa.int/web/planck/pla}} 

Poly-reion with respect to the hyperbolic tangent allows for a much greater freedom and encompasses from near-instantaneous to very extended reionization histories, where we define the duration of reionization as the difference between the redshifts where reionization was 10\% and 99\% done.This freedom translates in a general preference for higher optical depths with respect to the hyperbolic tangent case, although always in agreement inside the error bars, due to the non-instantaneous duration of reionization allowed by the model.

\section{Constraints on the history of reionization with Planck 2018}

The section is dedicated to the constraints on the Poly-reion model for reionization with Planck 2018 data in the context of the $\Lambda$CDM model.

\subsection{Planck 2018 baseline}

We start by considering the different combinations of the Planck 2018 baseline likelihood~\citep{Aghanim:2019ame}.
This choice of likelihoods contains a part at low-$\ell$, i.e. $2 \le \ell \le 29$, whose temperature part
uses a Gibbs sampler on a foreground cleaned combination of $30-353$ GHz channels obtained by the \texttt{commander} algorithm and
whose $E$-mode polarization is based on the 100x143 GHz cross angular power spectra.
At high multipoles, the Plik TTTEEE includes
auto and cross-spectra in temperature and polarization of
100, 143 and 217 GHz channels. The Planck 2018 baseline likelihood contains also the CMB lensing likelihood~\citep{Aghanim:2018oex}, in the conservative
range $8 \le \ell \le 400$. For more details on these likelihoos see \cite{Aghanim:2019ame,Aghanim:2018oex}.

We briefly summarize the improments of the Planck 2018 data with respect to the 2015 release.
The one that mostly affects the reionization history constraints is the change
in the large angular scale polarization~\cite{Aghanim:2019ame}. In Planck 2015 the baseline choice
was provided by a pixel based TQU likelihood, whose polarization part was provided
the Low Frequency Instrument (LFI) 70GHz channel cleaned by synchrotron and dust contamination with 30 GHz and 353 GHz, respectively.
In 2018 the lowE likelihood is provided by a simulation based likelihood that uses the High Frequency Instrument (HFI) 100x143 GHz cross-spectra, significantly
reducing the error bars and updating the 2016 first results \cite{Aghanim:2016yuo}. Other important differences are the use of only
$Planck$ data in the Commander map for the 2018 Gibbs low-$\ell$ likelihood and at high multipoles temperature to polarization leakage corrections,
that has validated the use of high-$\ell$ polarization and temperature-polarization cross-correlation for cosmological results.

\begin{figure}[!htb]
\includegraphics[width=\columnwidth]{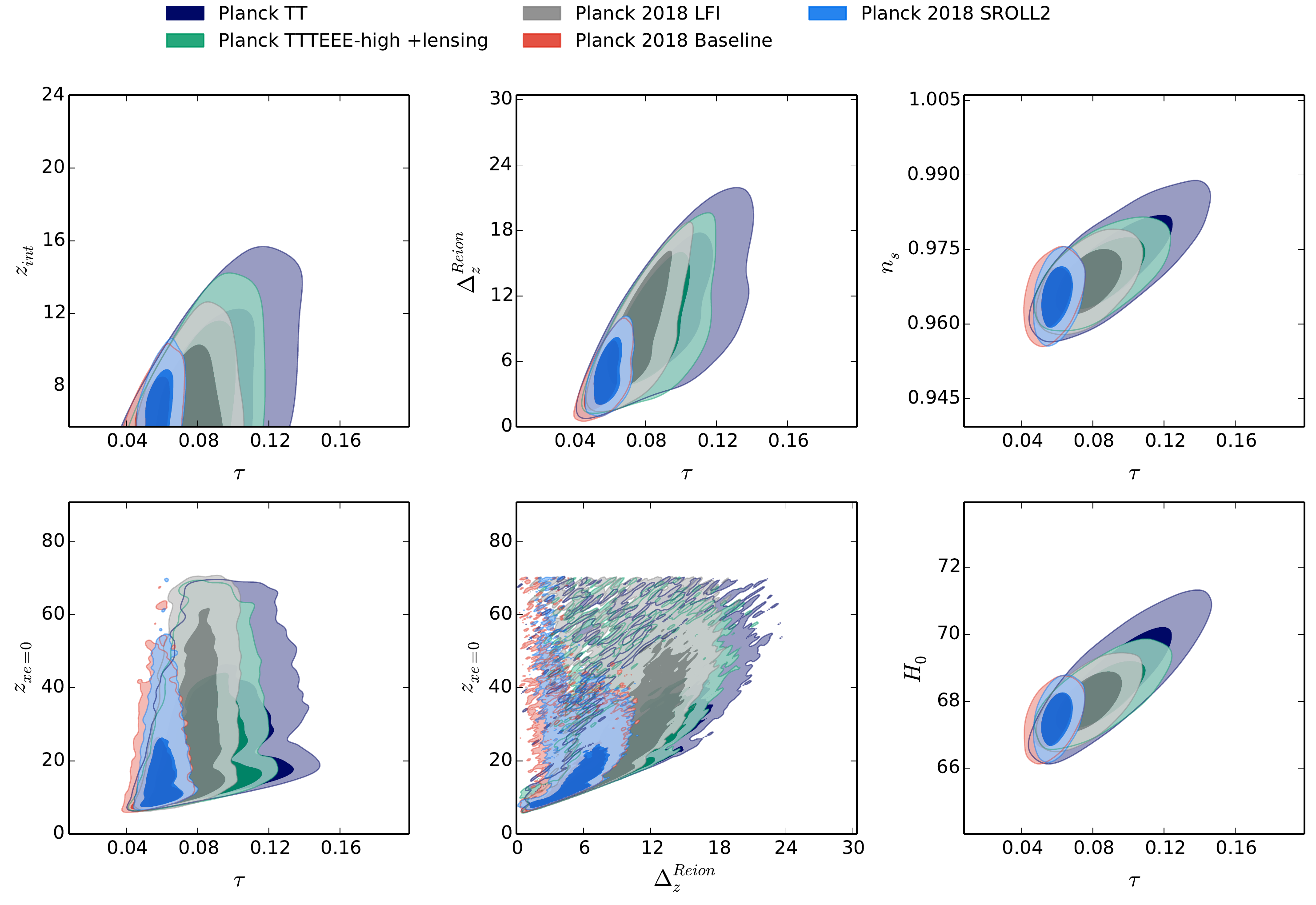}
\caption{\footnotesize\label{2D2018} 
 Marginalized joint 68\% and 95\% CL regions for cosmological and reionization parameters for different $Planck$ 2018 likelihoods.
}
\end{figure}

 The results from the Planck TTTEEE+lowE+lensing are shown in red in Fig.~\ref{2D2018} and presented in the first column of Table~\ref{Table1}.
The results are consistent with the overall trend shown also by other extended models of reionization: with 
HFI low-$\ell$ polarization the duration of reionization allowed by Planck 2015 is strongly reduced \cite{Adam:2016hgk,Millea:2018bko,Aghanim:2018eyx,Hazra:2019wdn}. With current data the Poly-reion model is consistent with a shorter duration of reionization given by  
$\Delta_z^{Reion}=4.6_{-2.5}^{+1.7}$ at 68\% CL and excludes very early onsets $z_{xe=0}=18.2_{-10.9}^{+1.7}$ at 68\% CL. 
The optical depth is in agreement with the hyperbolic tangent results with the tendency already shown in~\citep{HPFS18} for preferring higher central values $\tau=0.0572_{-0.0075}^{+0.0064}$ at 68\% CL,  compared to the Planck result $\tau=0.0544 \pm 0.0073$ \cite{Aghanim:2018eyx}. Overall we note the significant improvements on the constraints on Poly-reion with respect to 2015 data \cite{HPFS18} and the excellent consistency of the estimate of the remaining cosmological parameters with the standard use of the hyperbolic tangent case.
\begin{table}[!htb]
\centering
\begin{tabular}{|l|p{32mm}|p{32mm}|p{32mm}|}
\hline
Parameter/Data &Planck TTTEEE + lowE+ lensing & Planck TT+lensing  &Planck TTTEEE +lensing   \\
\hline
$\Omega_b$ & $0.0224\pm0.0001$& $0.0224\pm 0.0002$ & $ 0.0225\pm 0.0002$ \\
$\Omega_c$  & $0.120\pm0.001$& $0.117\pm0.002$ & $ 0.118\pm0.001$\\
$100\theta_{MC}$ & $1.0409\pm0.0003$& $1.0412\pm0.0005$ & $ 1.0411\pm0.0003$ \\
$\tau$  & $0.0572_{-0.0075}^{+0.0064}$& $0.0926_{-0.0232}^{+0.0196}$ & $ 0.0851_{-0.0157}^{+0.0158}$\\
${\mathrm{ln}}(10^{10}A_s)$  & $3.05\pm0.01$& $3.11\pm0.04$ & $ 3.10\pm0.03$\\
$n_s$  & $0.9656\pm0.0041$& $0.9723\pm0.0065$ & $ 0.9702\pm0.0048$\\
$H_0$  & $67.46\pm0.54$& $68.67_{-1.12}^{+0.99}$ & $ 68.21\pm0.66$\\
$z_{\mathrm re}$ & $7.24_{-1.11}^{+0.74}$ & $9.40_{-2.81}^{+1.69}$ & $ 8.90_{-2.28}^{+1.55}$\\
$z_{xe=0}$  & $18.18_{-10.89}^{+1.61}$& $28.24_{-18.28}^{+6.14}$ & $ 26.97_{-17.62}^{+5.65}$\\
$\Delta^{Reion}_{z}$ & $4.59_{-2.45}^{+1.67}$ & $10.47_{-5.13}^{+3.92}$ & $ 9.43_{-4.44}^{+3.57}$\\
\hline
\end{tabular}
\caption{\footnotesize\label{Table1} Confidence limits (68 \%) for cosmological parameters with different Planck 2018 likelihood combinations.
}
\end{table}

\begin{figure}[!htb]
\includegraphics[width=\columnwidth]{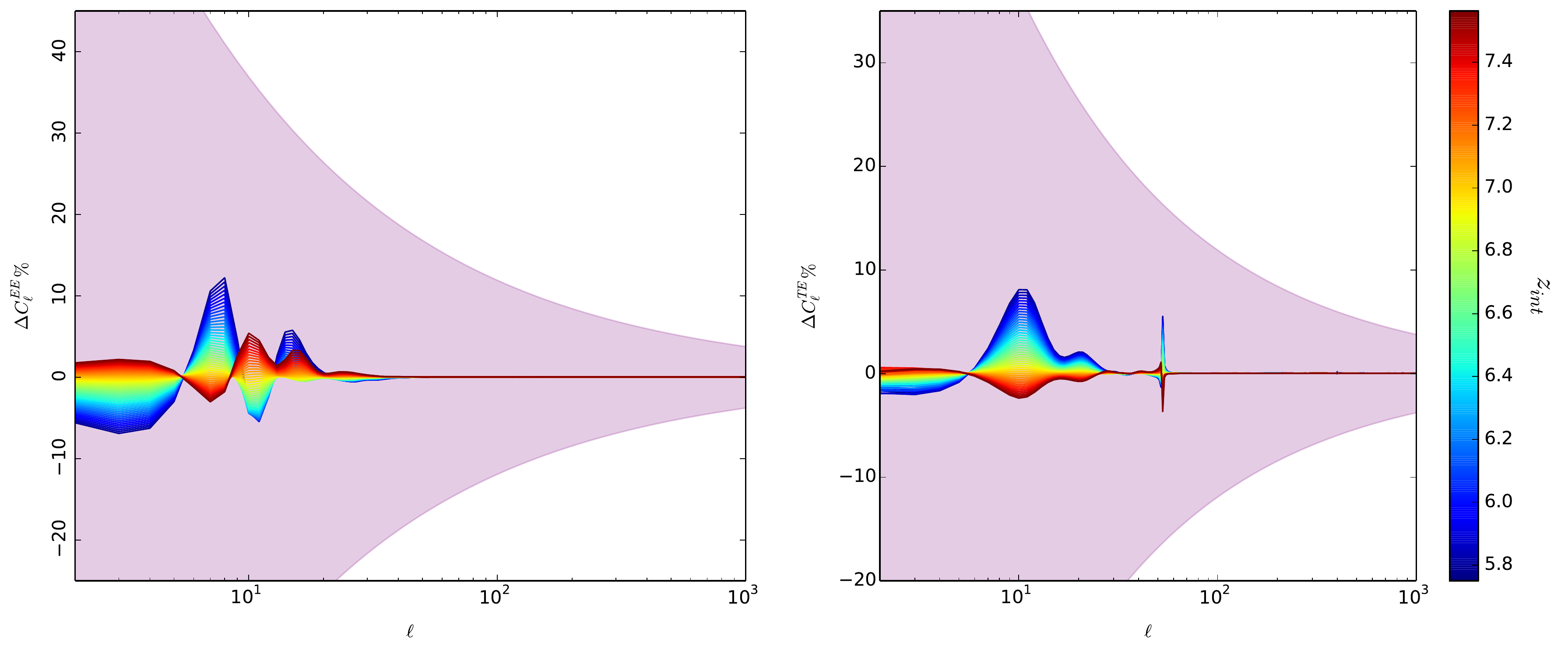}\\
\includegraphics[width=\columnwidth]{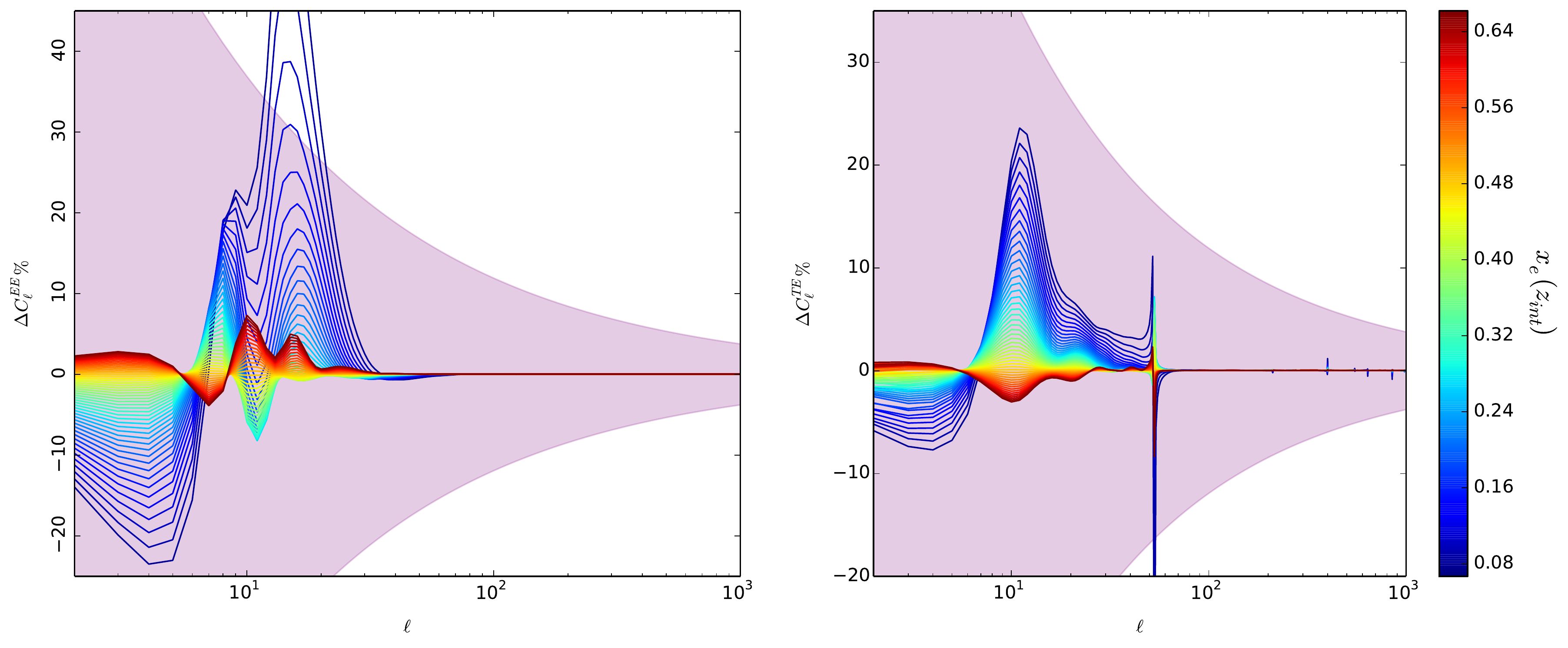}\\
\includegraphics[width=\columnwidth]{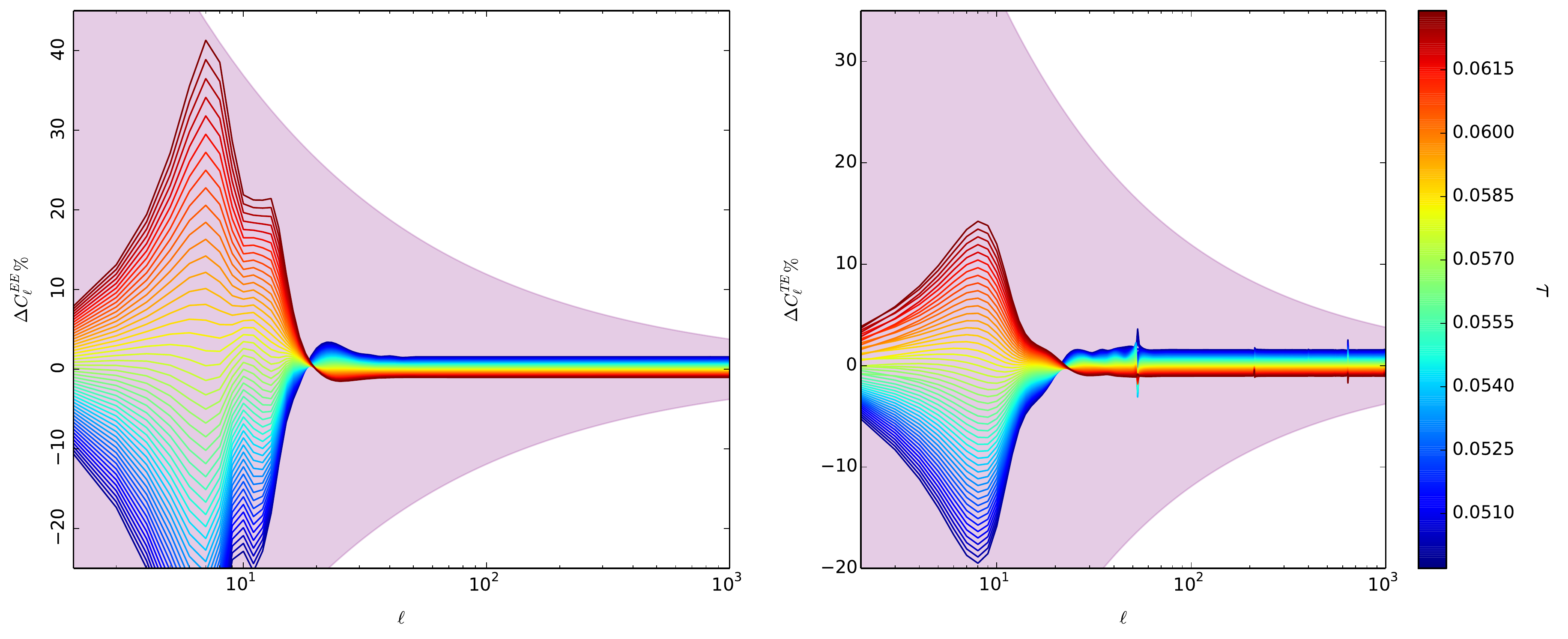}
\caption{\footnotesize\label{clvar}Variation of the E-mode polarization and TE cross-correlation angular power spectra with respect to the three Poly-reion parameters. Top panel is the position of the intermediate redshift, second panel is the amplitude the bottom panel is the variation with the optical depth.}
\end{figure}

In order to fully understand our results, we investigate the variation of the angular power spectra in polarization and TE cross correlation with the Poly-reion parameters. 
In particular we study the variation with respect to the best fit of our results varying the three main parameters within 1 sigma of the data constraints: the position of the intermediate point $z_{int}$ between [5.75,7.60] , the amplitude of the ionization fraction in the the point $xe_{z=int}$  between [0.0665,0.674], the optical depth $\tau$ between  [0.0497,0.0632].
In Fig \ref{clvar} we show the variation compared with the cosmic variance for a Planck-like experiment (70\% of the sky coverage). The varied parameter is shown in the colormap, all the other parameters are fixed to their best fit value.\\

In order to test the different pulls data have on the Poly-reion model and the robustness of the results, we have considered different combinations of Planck baseline data subsets.
We start by excising the large angular scale polarization from the likelihood combination. We consider only temperature data and use the leverage of the lensing to constrain the optical depth without imposing a Gaussian prior. The results are shown in purple contours in Fig.~\ref{2D2018} and presented in the second column of Table~\ref{Table1}. We first note the good agreement of the temperature plus lensing results with the baseline, though with an higher central value the optical depth is in agreement within 2 sigma.
The addition of high-ell polarization tightens the error bars and pushes the optical depth towards lower values as shown by the green contours in Fig.~\ref{2DLCDM} and  the third column of Table~\ref{Table1}.

\subsection{Alternatives to Planck 2018 baseline}

We test now the two available alternative likelihoods to the Planck 2018 baseline for large scale polarization. 
In the 2018 release together with the LowE likelihood it has been released a full TQU pixel based likelihood from the LFI data at 70 GHz (plus the same temperature map as low$l$), lowTEB (note that a joint likelihood LFI-WMAP has been recently developed in \cite{Natale:2020owc}). This alternative likelihood is relevant, although it produces larger error bars, because it considers all the channels, TT, EE, BB and also TE, where the latter is not present in the baseline lowE, and it represents both a change in the data and a change in the likelihood algorithm, providing therefore an optimal robustness test. The results of this alternative likelihood are represented by the grey contours in Fig.~\ref{2D2018}. Considering the larger error bars the optical depth $\tau=0.078 \pm 0.012$ at 68\% CL shows a very good agreement with the HFI based baseline, although with a slightly larger central value, a tendency shown also by the hyperbolic tangent.

We also test the alternative re-processing of Planck HFI data based on the SROLL2 map making algorithm~\cite{Delouis:2019bub} which reports an improvement in the removal of systematics with respect to the baseline Planck 2018 data analysis. The $\Lambda$CDM+ Tanh optical depth given by this likelihood, without the lensing addition, is $\tau=0.0590^{+0.0058}_{-0.0068}$ at 68\% CL~\cite{Pagano:2019tci} with a slightly larger central value with respect to the Planck 2018 baseline result. We confirm this tendency towards higher values also with the Poly-reion model where we obtain $\tau=0.0600^{+0.0052}_{-0.0064}$. The result is shown in blue in Fig.~\ref{2D2018} and is in good agreement with the hyperbolic tangent case though with a slightly improved error bars given by the addition of the lensing likelihood which is not present in the hyperbolic tangent result.

\vspace{1cm}

\begin{figure}[!htb]
\includegraphics[width=\columnwidth]{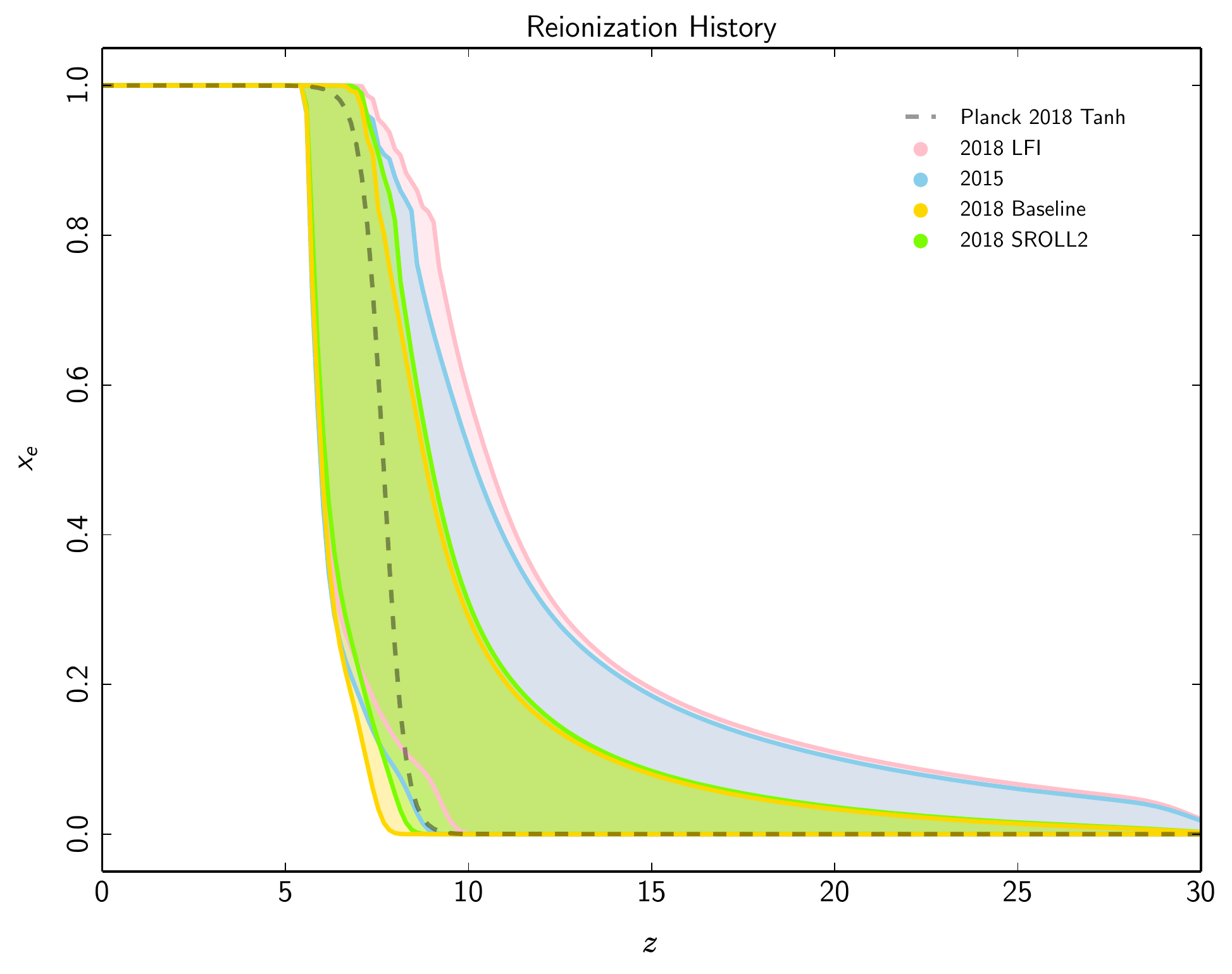}
\caption{\footnotesize\label{xe} 
Derived 95\% confidence limits for $x_e (z)$ by the Poly-reion model using different Planck 2018 likelihood combinations. The best-fit for $x_e (z)$ from the Planck 2018 baseline likelihoods in the hyperbolic tangent case is displayed as a dashed line for comparison.} 
\end{figure}

\begin{figure}[!htb]
\includegraphics[width=\columnwidth]{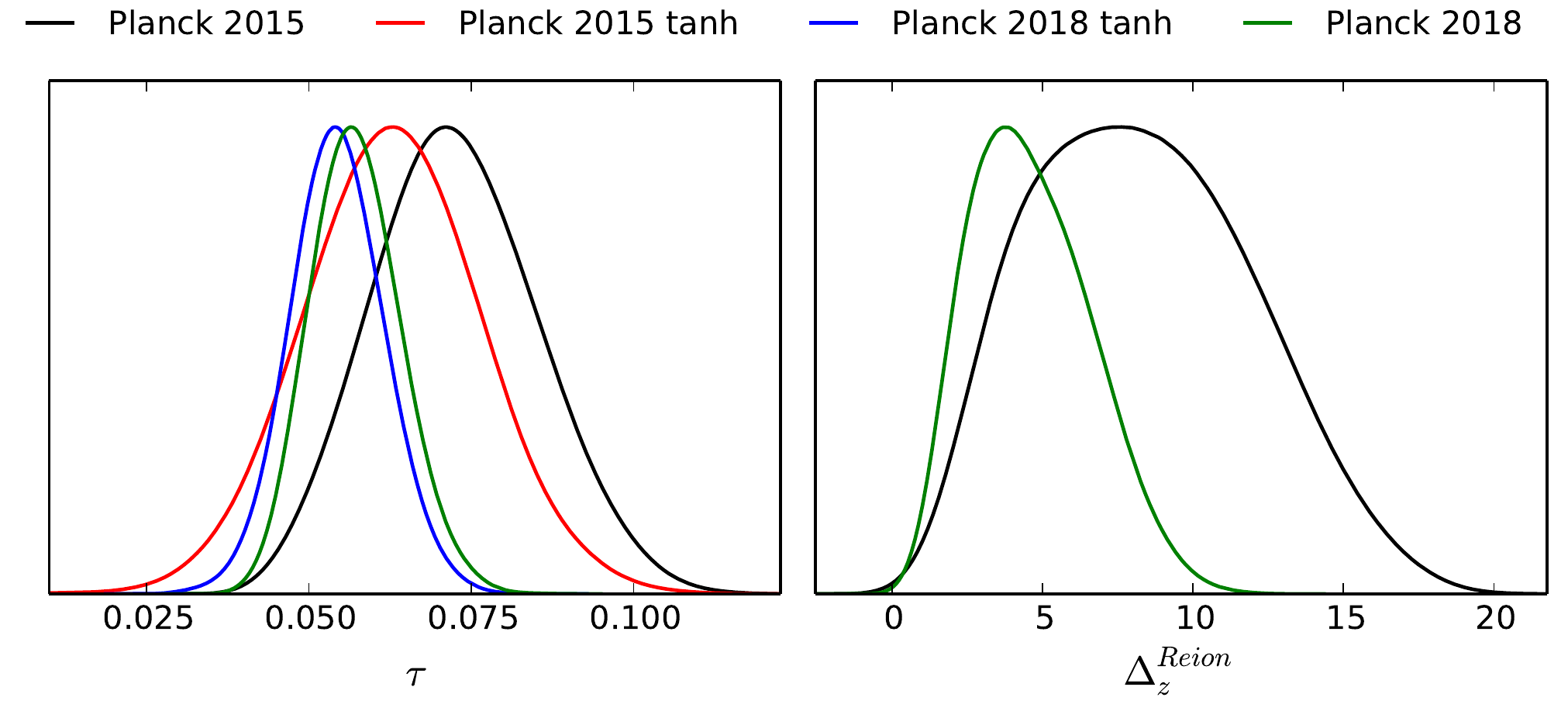}
\caption{\footnotesize\label{1DLCDM}  Marginalized probability density for $\tau$ and $\Delta_z^{Reion}$ for different Planck data. The standard results for $\tau$ with Planck 2015 (red) and 2018 (blue) are shown for comparison.
}
\end{figure}

In summary, the Planck 2018 data show a large improvement in the constraints to the Poly-reion model thanks mainly to the improved polarization on large angular scales. Long duration of reionization and early onsets, hinted by Planck 2015 data, are now strongly disfavoured with  $\Delta_z^{Reion}=4.6_{-2.5}^{+1.7}$ at 68\% CL and  $z_{xe=0}=18.2_{-10.9}^{+1.7}$ at 68\% CL. The optical depth is in agreement with the hyperbolic tangent case with a slightly higher value $\tau=0.0572_{-0.0075}^{+0.0064}$ at 68\% CL. The results are consistent against subsets of Planck data and alternatives to the baseline demonstrating the robustness of the new Planck 2018 against different models of reionization.
Fig.~\ref{xe} and Fig.~\ref{1DLCDM} summarize the consistency and improvements of the new results.

In Fig.~\ref{xe} we present the ionization fraction history as constrained from the different data sub-combinations. We note how also for the ionization fraction evolution all the combinations are in good agreement, mirroring the results on the optical depth, with LFI-based results (both 2015 and 2018) allowing for earlier onsets of the reionization with respect to HFI-based data.    

The improvement of Planck 2018 baseline results with respect to the 2015 ones for the hyperbolic tangent is reflected also by a substantial improvement in the constraints on the Poly-reion. In particular, in  Fig.~\ref{1DLCDM} we show the comparison of the optical depth as constrained by Planck 2015 and Planck 2018 for the Poly-reion model compared with the analogous for the standard hyperbolic tangent. We note how the tendency to drive upwards the value of the optical depth in the Poly-reion model shown in 2015 data is much reduced with Planck 2018, thanks mainly to the improved polarization data. In the right panel of Fig.~\ref{1DLCDM} we show the comparison of the duration of reionization in Poly-reion between 2015 and 2018. We note that with Planck 2018 the duration of reionization is much tightly constrained ruling out the high redshift tails that were still allowed by Planck 2015 data.

\begin{figure}[!htb]
 \includegraphics[width=\columnwidth]{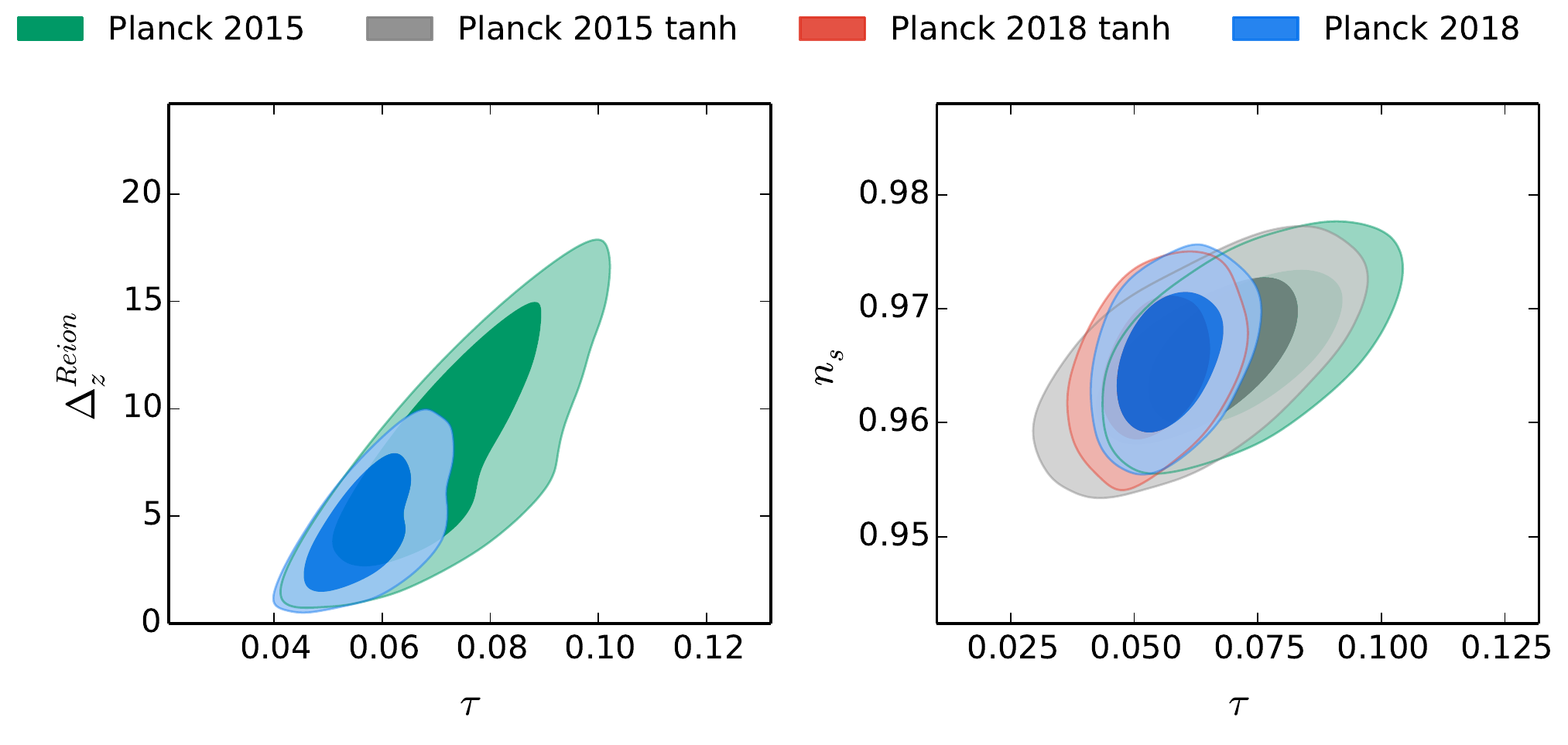}
\caption{\footnotesize\label{2DLCDM} Marginalized joint 68\% and 95\% CL regions for $(\tau,\Delta_z^{Reion})$ and $(\tau,n_s)$ with  different $Planck$ 2018 likelihoods.
}
\end{figure}

\begin{figure}[!htb]
\includegraphics[width=0.5\columnwidth]{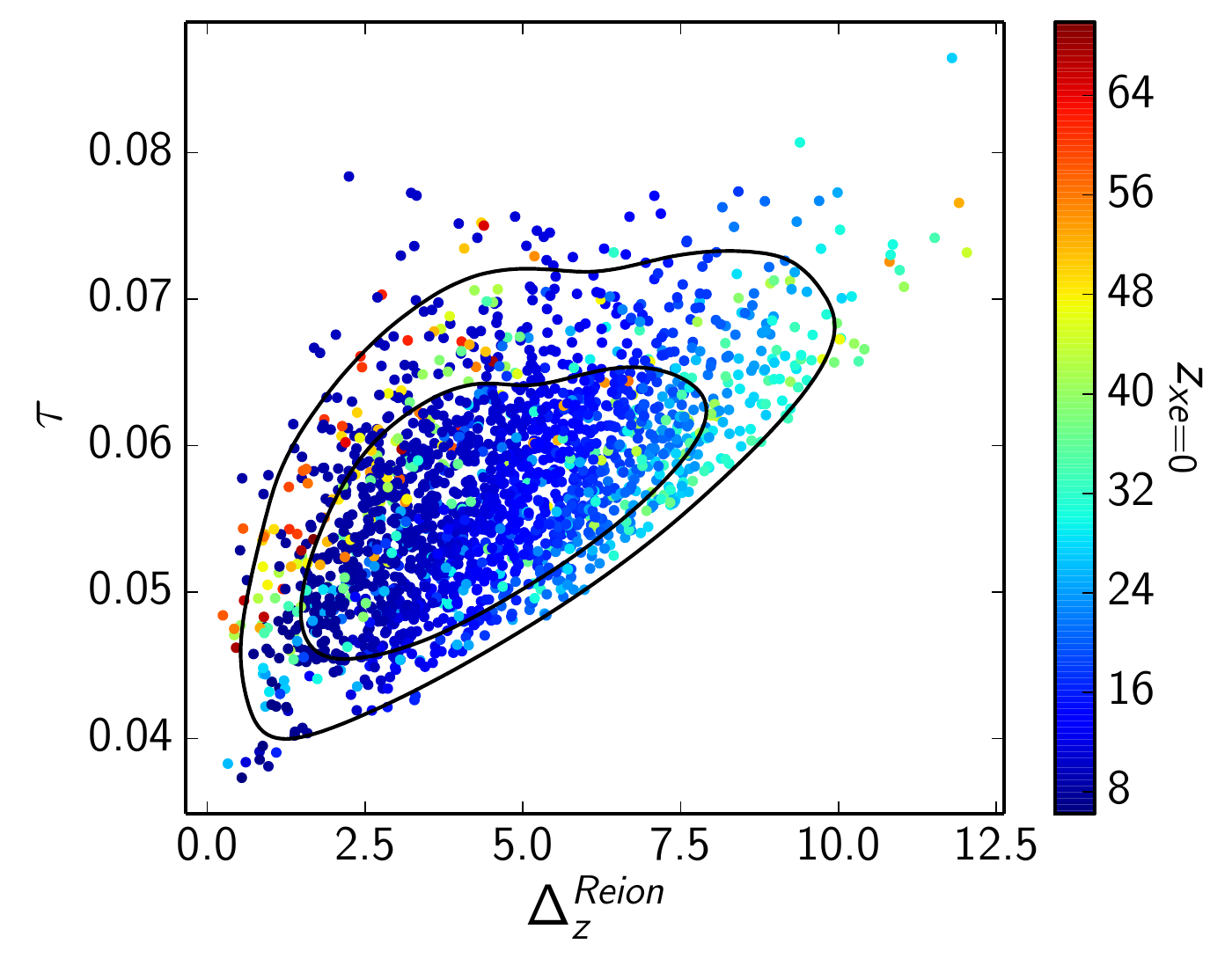}\includegraphics[width=0.5\columnwidth]{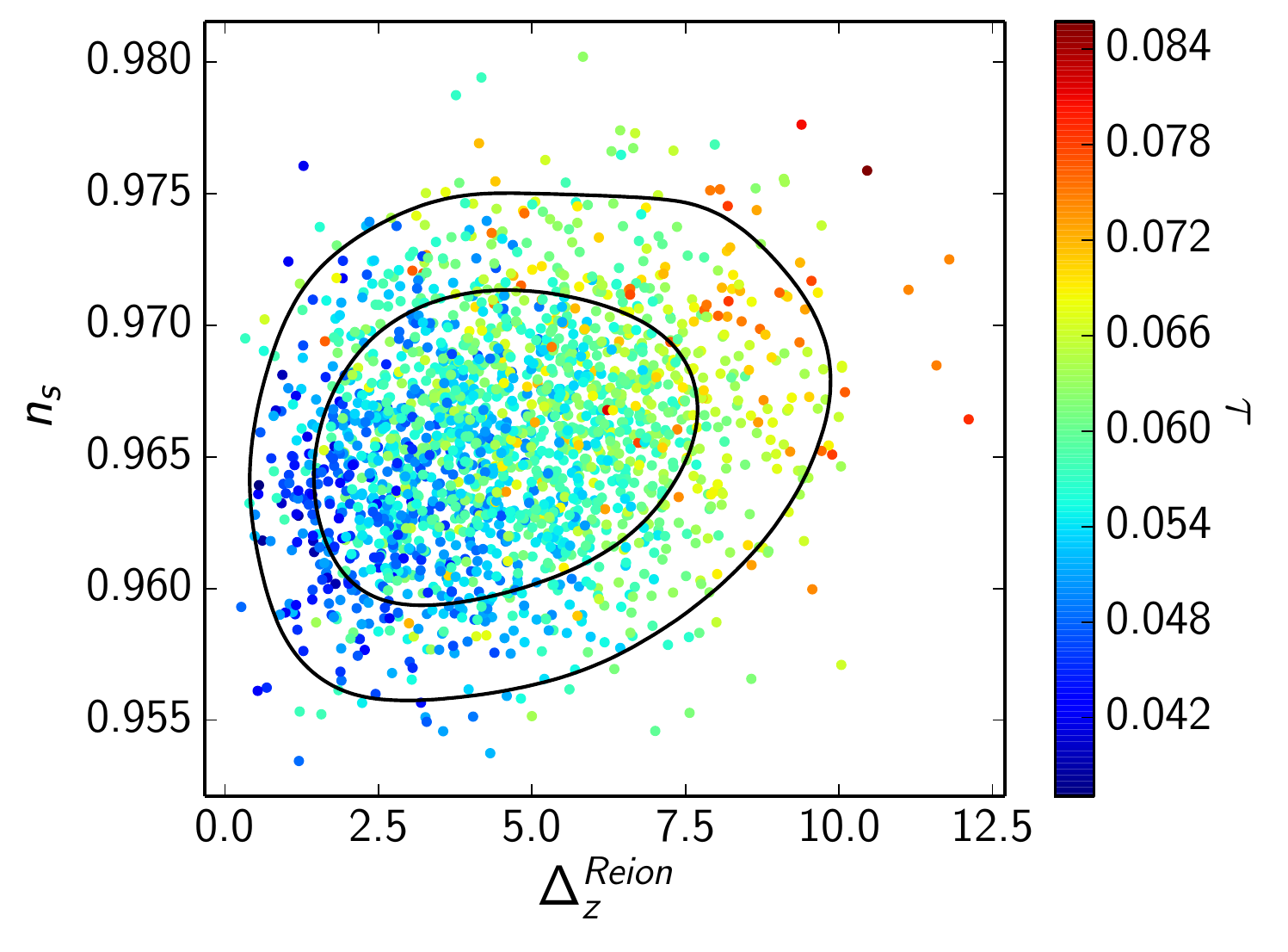}
\caption{\footnotesize\label{3D1}Three dimensional posterior distribution for Poly-reion parameters as the duration of reionization and the starting redshift compared with the optical depth and the scalar spectral index for the $\Lambda$CDM model. }
\end{figure}

In Fig.~\ref{2DLCDM} we show the breaking of the optical depth versus the scalar spectral index degeneracy, already observed for the hyperbolic tangent, is present also in the Poly-reion model thanks also the tightening in the $\tau$ direction. We notice also a drastic change in the scatter plot in Fig.~\ref{3D1} with respect to the analogous of~\cite{HPFS18} and we especially note again how early onsets are now disfavoured in Poly-reion with Planck 2018 data.

\newpage
\section{Extensions to the $\Lambda$CDM}
We now investigate the constraints of Planck 2018 baseline data on Poly-reion when an extended cosmological model is assumed. 
We focus on the extended sectors which have known degeneracies with the reionization model updating our previous results with Planck 2015 ~\cite{HPFS18} and adding new cases, such as spatial curvature, number of effective neutrinos and its combination with the sum of neutrino masses, a large scale power suppression and the lensing amplitude. 

\subsection{$\Lambda$CDM+tensor}
We start our compilation of extensions to $\Lambda$CDM model by adding a primordial tensor contribution. In this case the main impact is on large angular scales and we expect a slight degeneracy between Poly-reion and the amplitude of tensor modes. We let the tensor to scalar ratio $r$ free to vary together with the standard cosmological parameters and the Poly-reion ones.
\begin{figure}[!htb]
\includegraphics[width=\columnwidth]{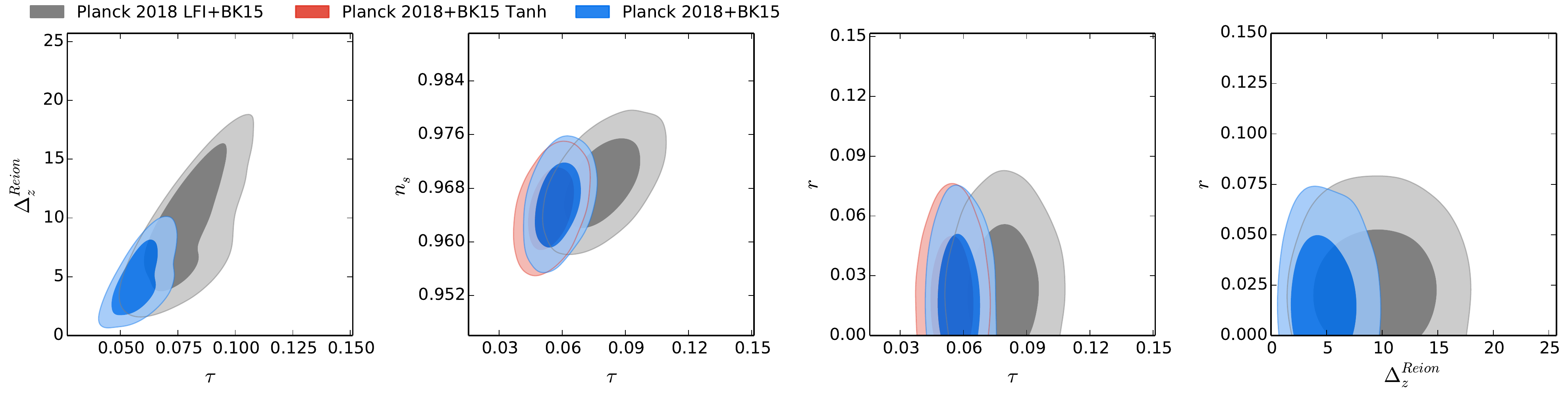}
\caption{\footnotesize\label{2D2} Marginalized joint 68\% and 95\% CL regions for $(\tau,\Delta_z^{Reion})$, $(\tau,n_s)$,$(\tau,r)$ and $(r,\Delta_z^{Reion})$ compared with the hyperbolic tangent case.}
\end{figure}

To add constraining power on the tensor to scalar ratio to the Planck data we combine the Planck 2018 baseline with the Bicep/Keck data on B-mode polarization from 2015 observations\citep{BK15}.
As robustness test we also consider the alternative LFI based Planck 2018 likelihood previously described.

In Table~\ref{Table2} we present the results and in Fig.~\ref{2D2} we show the two dimensional constraints on the reionization parameters and r compared with the analogous case with the hyperbolic tangent model for reionization. We note that the introduction of the Poly-reion model does not impact the constraints on the tensor to scalar ratio, that are in perfect agreement with the standard case \cite{Akrami:2018odb}. The two-dimensional posterior distributions  do not show any significant degeneracy between the Poly-reion parameters and the tensor to scalar ratio. The use of HFI-based polarization significantly reduces the allowed duration of reionization and delays its onsets with respect to the LFI-based likelihood in the same manner as for the $\Lambda$CDM case.
We can conclude that the addition of a primordial tensor contribution does not affect either the constraints on primordial tensor amplitudes nor the constraints on the model of reionization.
\begin{table}[!htb]
\centering
\begin{tabular}{|l|l|l|l|}
\hline
Parameter/Data& Planck TTTEEE+lowE+lensing+BK15  &Planck TTTEEE +LFI+lensing+BK15 \\
\hline
$\Omega_b$ & $0.0224\pm 0.0001$ & $ 0.0225\pm 0.0002$\\
$\Omega_c$ & $0.120\pm 0.001$ & $ 0.119\pm 0.001$\\
$100\theta_{MC}$ & $1.0409\pm 0.0003$ & $ 1.0410\pm 0.0003$\\
$\tau$ & $0.0581_{-0.0077}^{+0.0065}$ & $ 0.0793_{-0.0120}^{+0.0119}$\\
${\mathrm{ln}}(10^{10}A_s)$ & $3.05\pm 0.01$ & $ 3.09\pm 0.02$\\
$n_s$ & $0.9657_{-0.0042}^{+0.0041}$ & $ 0.9689\pm 0.0044$\\
$r$ & $<0.07$ & $ <0.06$\\
$H_0$ & $67.40\pm0.54$ & $ 67.97\pm0.60$\\
$z_{\mathrm re}$ & $7.28_{-1.14}^{+0.77}$ & $ 7.95_{-2.15}^{+0.83}$\\
$z_{xe=0}$ &$ 18.82_{-11.36}^{+1.82}$ & $32.13_{-22.98}^{+8.71}$  \\
$\Delta^{Reion}_{z}$ &$ 4.80_{-2.47}^{+1.71}$ &  $9.52_{-4.05}^{+3.98}$\\
\hline
\end{tabular}
\caption{\footnotesize\label{Table2} Confidence limits at 68 \% (95\% for upper limits) for cosmological parameters in the $\Lambda$CDM+r model.}
\end{table}

\subsection{Neutrino Physics}
We now study how the extension to Poly-reion model for reionization affects the knowledge on neutrino physics from the CMB data.

\subsubsection{$\Lambda$CDM+Mnu}
We start by assuming a non vanishing neutrino mass. Being Planck data not sensitive to the hierarchy choice, we consider a degenerate hierarchy . 
We vary the neutrino mass jointly with cosmological and Poly-reion parameters.
\begin{figure}[!htb]
\includegraphics[width=\columnwidth]{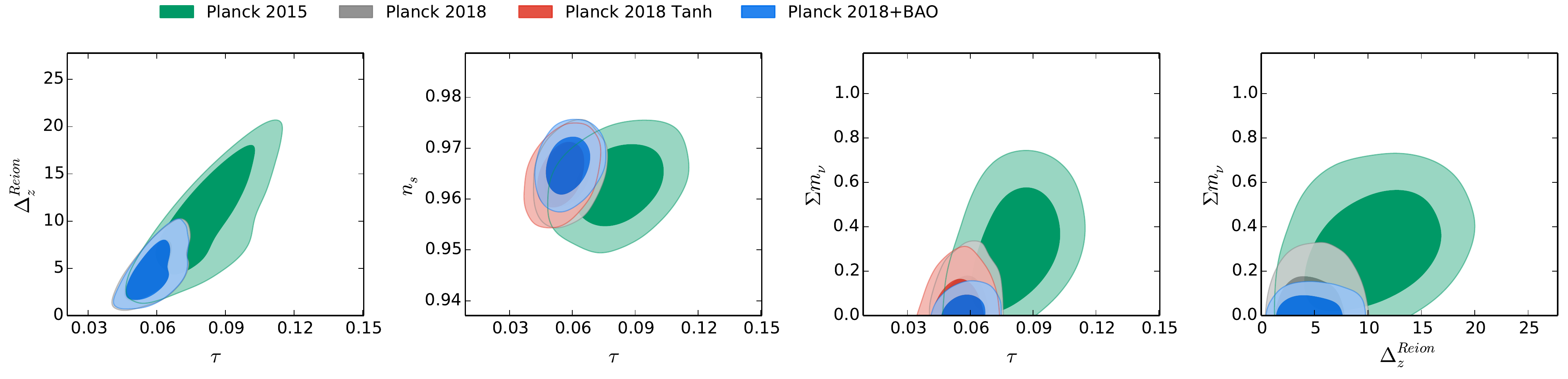}
\caption{\footnotesize\label{2Dmnu}Marginalized joint 68\% and 95\% CL regions for $(\tau,\Delta_z^{Reion})$, $(\tau,n_s)$,$(\tau,\Sigma m_\nu)$ and $(\Sigma m_\nu,\Delta_z^{Reion})$ compared with the hyperbolic tangent case and the Planck 2015 results. We also consider the case of the BAO addition in blue.}
\end{figure}
\begin{figure}[!htb]
\includegraphics[width=0.45\columnwidth]{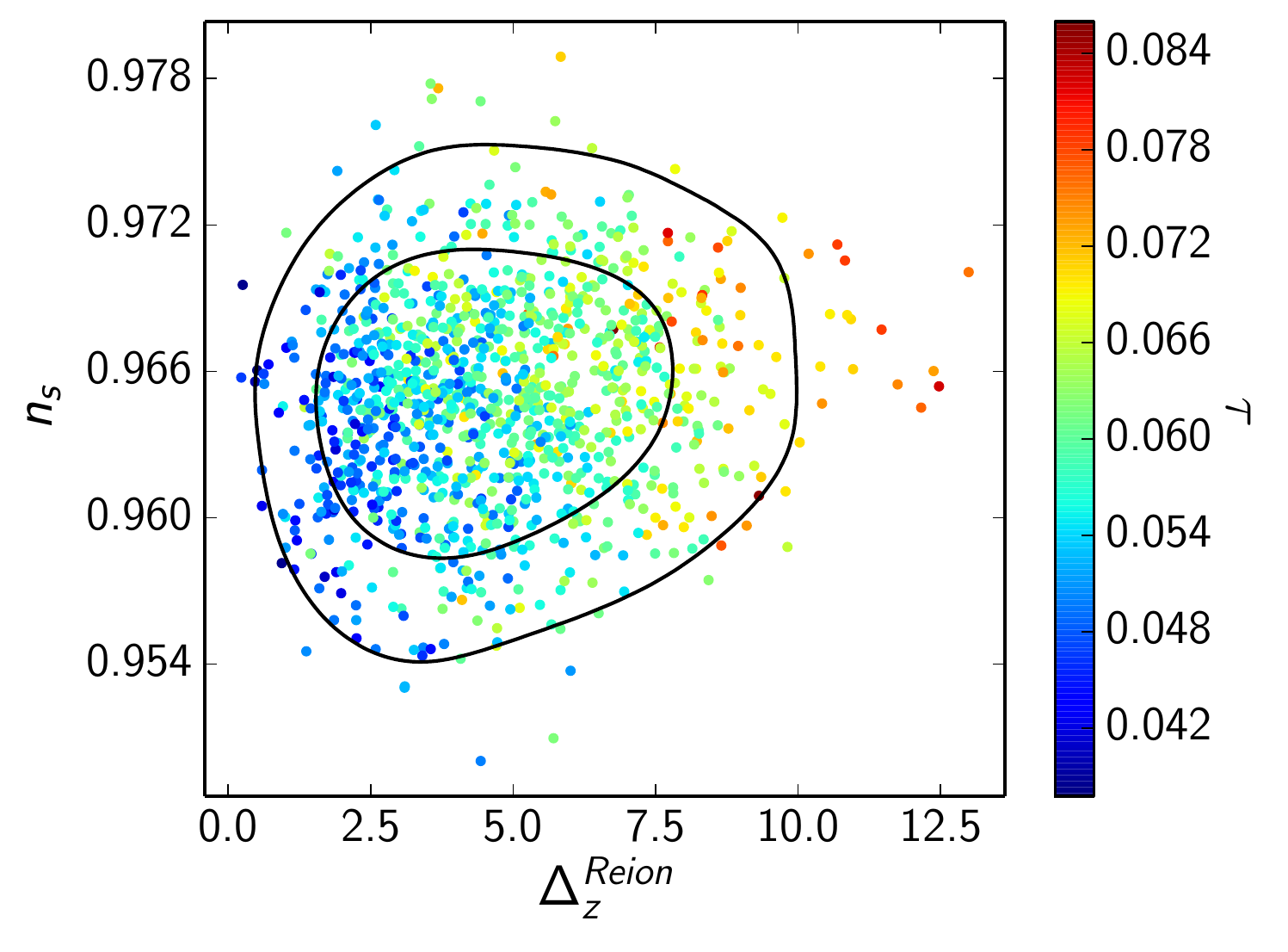}\includegraphics[width=0.45\columnwidth]{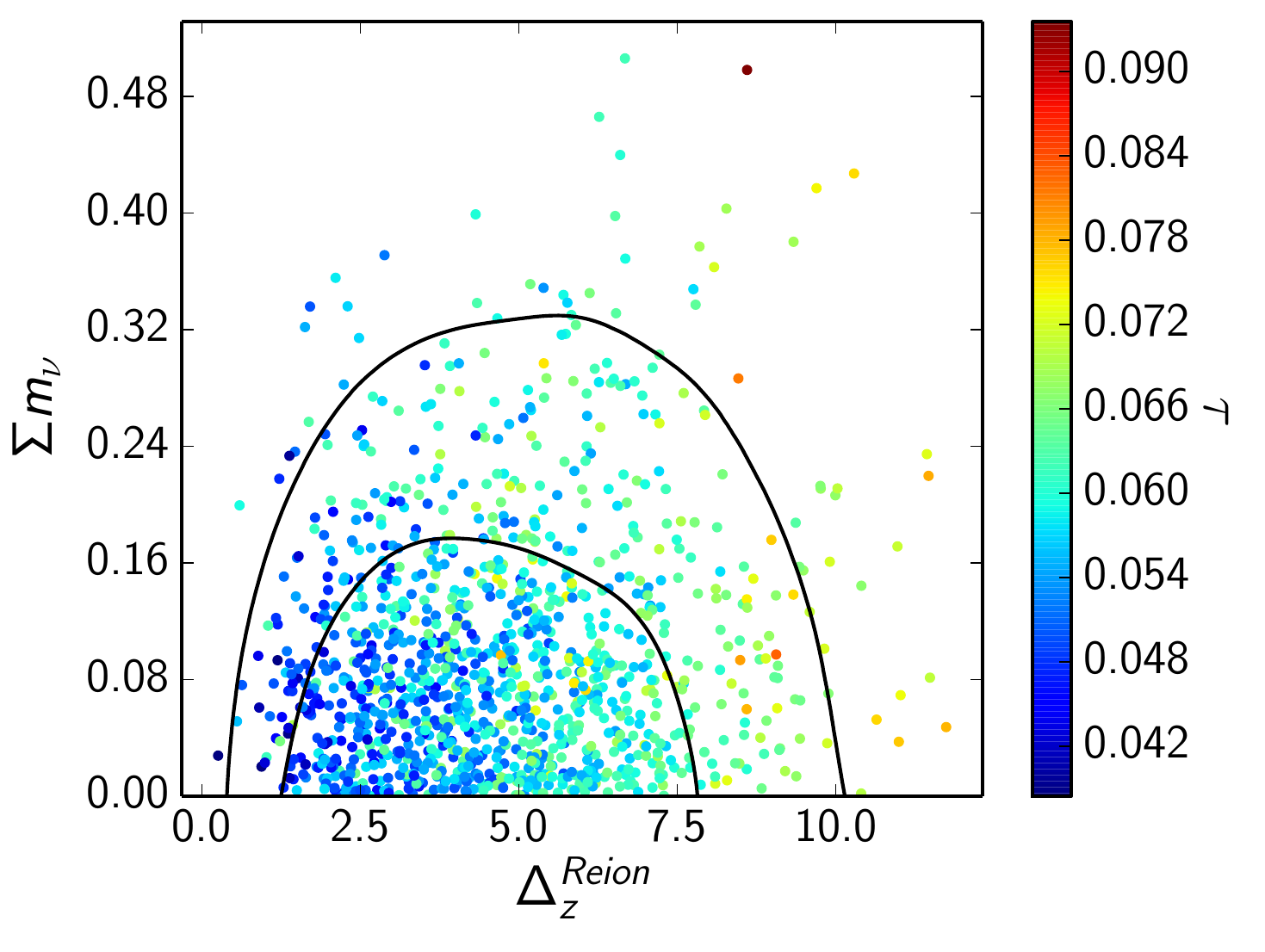}
\caption{\footnotesize\label{3D3}Three dimensional marginalized posteriors for the duration of reionization versus the scalar spectral index and the sum of neutrino masses. The color scale is referred to the optical depth.}
\end{figure}
The resulting two dimensional distributions for the Poly-reion parameters and the neutrino mass are shown in Fig.~\ref{2Dmnu}. The resulting constraints on the neutrino mass are $M_\nu<0.266$ eV at 95\%CL, reduced to $M_\nu<0.126$ eV at 95\%CL when the Baryonic Acoustic Oscillations data are added to the analysis. The constraints are compatible with the hyperbolic tangent ones, respectively $M_\nu<0.24$ eV at 95\%CL and $M_\nu<0.12$ eV at 95\%CL \cite{Aghanim:2018eyx}. The minor increase in the upper limit is given by the larger optical depth in Poly-reion, that for the baseline data case is  $\tau=0.0580_{-0.0080}^{+0.0067}$ at 68\%CL with $z_{xe=0}=17.92_{-10.47}^{+1.47}$  at 68\%CL and $\Delta^{Reion}_{z}=4.71_{-2.51}^{+1.60}$  at 68\%CL with only a marginal improvement for the reionization history by the addition of the BAO.
The comparison in Fig.~\ref{2Dmnu} between grey, Planck 2018,  and green, Planck 2015, contours shows the amazing improvements of Planck 2018 data for this case. In 2015 the strong degeneracy between the neutrino mass and reionization was significantly degrading both constraints on the neutrino sector and reionization. With Planck 2018 data the degeneracy is almost completely broken allowing constraints at the level of the standard hyperbolic tangent.The improvement is also shown in the scatter plot in Fig.~\ref{3D3} where we note the very tight color scale for the optical depth.

\subsubsection{$\Lambda$CDM+$N_{\mathrm eff}$}
\begin{figure}[!htb]
\includegraphics[width=\columnwidth]{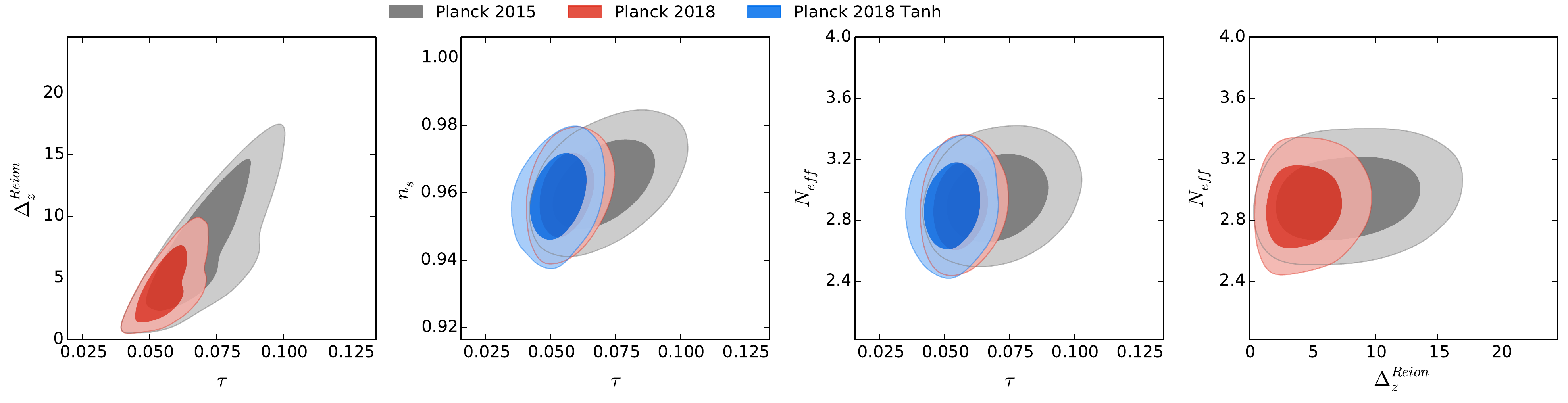}
\caption{\footnotesize\label{2Dnnu} Marginalized joint 68\% and 95\% CL regions for $(\tau,\Delta_z^{Reion})$, $(\tau,n_s)$,$(\tau,N_{eff})$ and $(N_{eff},\Delta_z^{Reion})$ compared with the hyperbolic tangent case and the 2015 results.}
\end{figure}

Another degeneracy affecting the neutrino sector and the reionization model is given by the one concerning the additional relativistic degrees of freedom, parametrized by varying the number of effective neutrinos. The grey contours in Fig.~\ref{2Dnnu} shows the constraints of Planck 2015 where the degeneracy is neat.
We now test if the degeneracy holds also with Planck 2018 data, the results are presented by the red contours in Fig.~\ref{2Dnnu} were we can already note significant improvement brought by the new Planck data. The degeneracy is broken and the constraints on the number of effective neutrinos are $N_{\mathrm eff}=2.90 \pm 0.19$ at 68\%CL in agreement with $N_{\mathrm eff}=2.89_{-0.18}^{+0.19}$ at 68\%CL from the hyperbolic tangent \cite{Aghanim:2018eyx}. The optical depth is $\tau=0.0564_{-0.0076}^{+0.0061}$ with $z_{xe=0}=17.87_{-10.72}^{+1.58}$ and $\Delta^{Reion}_{z}=4.43_{-2.47}^{+1.54}$.

\subsubsection{$\Lambda$CDM+Neff+Mnu}
Concluding the overview of the simplest neutrino sector extensions we consider the joint constraints on the number of effective neutrinos and the neutrino mass.
\begin{figure}[!htb]
\includegraphics[width=\columnwidth]{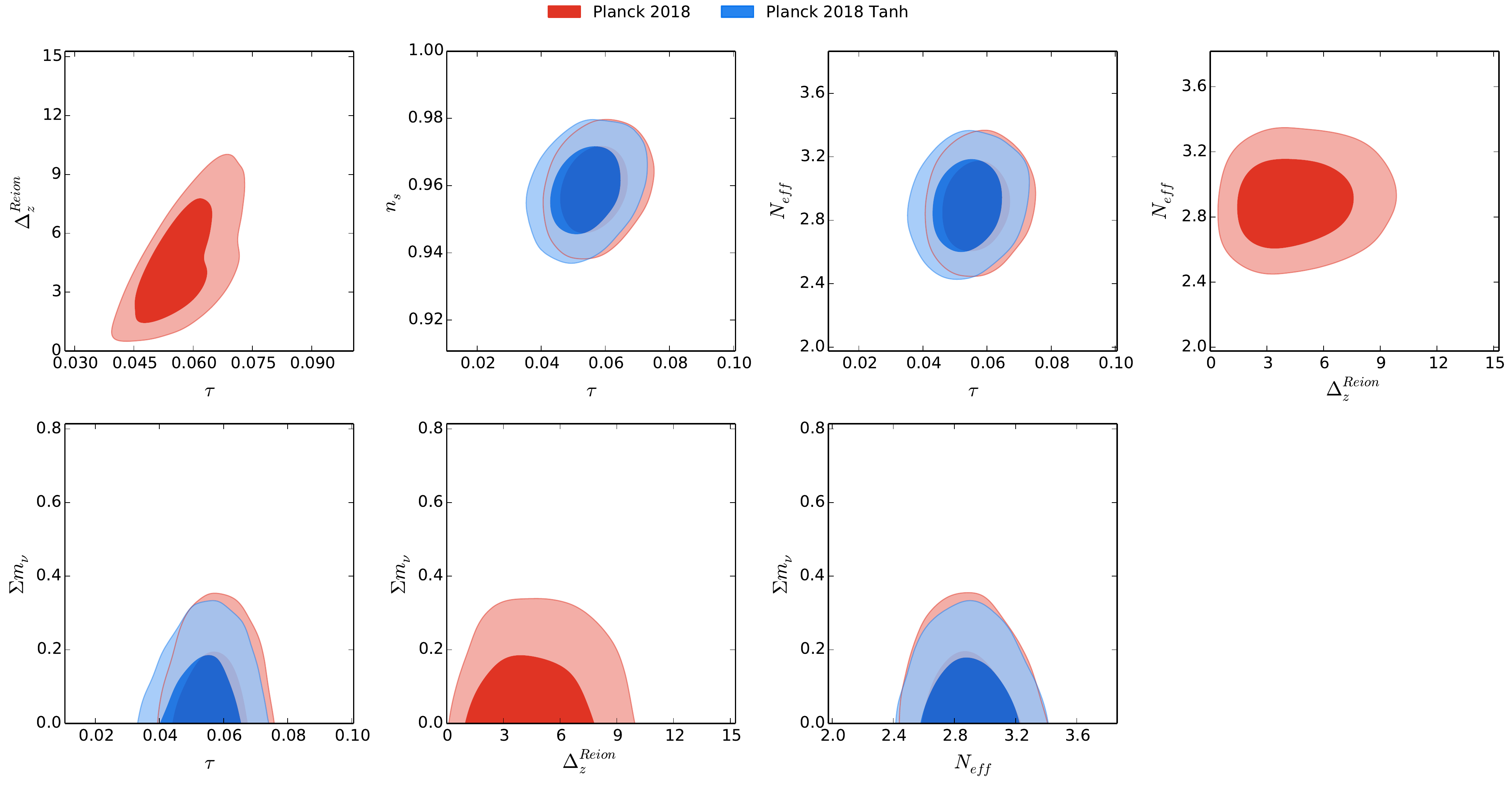}
\caption{\footnotesize\label{2Dmnunnu}Marginalized joint 68\% and 95\% CL regions for the neutrino sector and Poly-reion parameters compared with the hyperbolic tangent case.}
\end{figure}
In Fig.~\ref{2Dmnunnu} we show the results for the main parameters. We note that also in the case of joint constraints the data are sensitive enough to not suffer from degeneracies and provide constraints on the neutrino physics which are at the same level of the one provided by the standard hyperbolic tangent model: $N_{\mathrm eff}=2.89_{-0.20}^{+0.18}$ at 68\% CL and  $M_\nu<0.28$ to be compared with $N_{\mathrm eff}=2.90_{-0.19}^{+0.18}$ at 68\% CL and  $M_\nu<0.26$ at 95\% CL \cite{Aghanim:2018eyx}. 
Concerning the reionization part we note that in comparison with the hyperbolic tangent Poly-reion provides a constraint on $\tau$ with asymmetric uncertainties $\tau=0.0569_{-0.0076}^{+0.0063}$ to be compared with the hyperbolic tangent $\tau=0.0539_{-0.0072}^{+0.0071}$ as in previous models \cite{Aghanim:2018eyx}.
In Table~\ref{Table3} we summarize the results obtained for the physics of neutrinos.
\begin{table}[!htb]
\centering
\begin{tabular}{|l|l|l|l|}
\hline
Parameter/Model& $\Lambda$CDM+$\Sigma m_\nu$  & $\Lambda$CDM+$N_{\mathrm eff}$ & $\Lambda$CDM+$\Sigma m_\nu$+ $N_{\mathrm eff}$ \\
\hline
$\tau$ & $0.0580_{-0.0080}^{+0.0067}$ &$0.0564_{-0.0076}^{+0.0061}$ &  $0.0569_{-0.0076}^{+0.0063}$\\
$M_\nu$ & $<0.266$& - & $<0.28$ \\
$N_{eff}$ & - & $2.90 \pm 0.19$ & $2.89_{-0.20}^{+0.18}$ \\
${\mathrm{ln}}(10^{10}$ & $3.052_{-0.016}^{+0.014}$ & $3.042_{-0.017}^{+0.015}$&$3.043_{-0.018}^{+0.015}$\\
$n_s$ & $0.9650\pm 0.0042$ & $0.9596\pm 0.0081$ & $0.9590 \pm 0.0085$\\
$z_{\mathrm re}$ & $7.34_{-1.07}^{+0.80}$ & $7.15_{-1.08}^{+0.67}$& $7.19_{-1.13}^{+0.69}$\\
$z_{xe=0}$ & $17.92_{-10.47}^{+1.47}$&$17.87_{-10.72}^{+1.58}$&$18.27_{-11.00}^{+1.67}$ \\
$\Delta^{Reion}_{z}$ & $4.71_{-2.51}^{+1.60}$& $4.43_{-2.47}^{+1.54}$&$4.54_{-2.49}^{+1.61}$\\
\hline
\end{tabular}
\caption{\footnotesize\label{Table3} Confidence limits at 68\% ( 95\% for upper bounds) for neutrino sector parameters within the Poly-reion model.}
\end{table}
\subsection{$\Lambda$CDM+$A_{L}$}

We proceed by considering the case of the lensing amplitude $A_{L}$. We stress that $A_{L}$ is not a physical parameter, it phenomenologically represents the amplitude of the lensing effect on the CMB anisotropy angular power spectra but is not related to the physical lensing signal amplitude, that is measured through the four point correlation function on maps and that is at the base of the lensing likelihood~\citep{Aghanim:2018oex}. In order to study the correlation between the reionization history and the $A_{lens}$ without biases we do not include the lensing likelihood in the following analysis.

The reason beyond considering this specific extension of the $\Lambda$CDM is that although $A_{L}$ is not physical parameter it's a tracer of possible effects in the high multipoles region of the CMB angular power spectra and in particular of smoothing/oscillatory patterns in the acoustic peaks region. For this reason it can be degenerate with the effect of reionization on high multipoles, in particular when it's considered an extended cosmological model. We also remind that its value in the Planck 2018 baseline with the hyperbolic tangent is slightly higher than unity $A_{L}=1.180\pm0.065$ at 65\% CL indicating a preference of Planck data for an higher lensing smoothing than predicted~\cite{Aghanim:2018eyx}.   
\begin{figure}[!htb]
\includegraphics[width=\columnwidth]{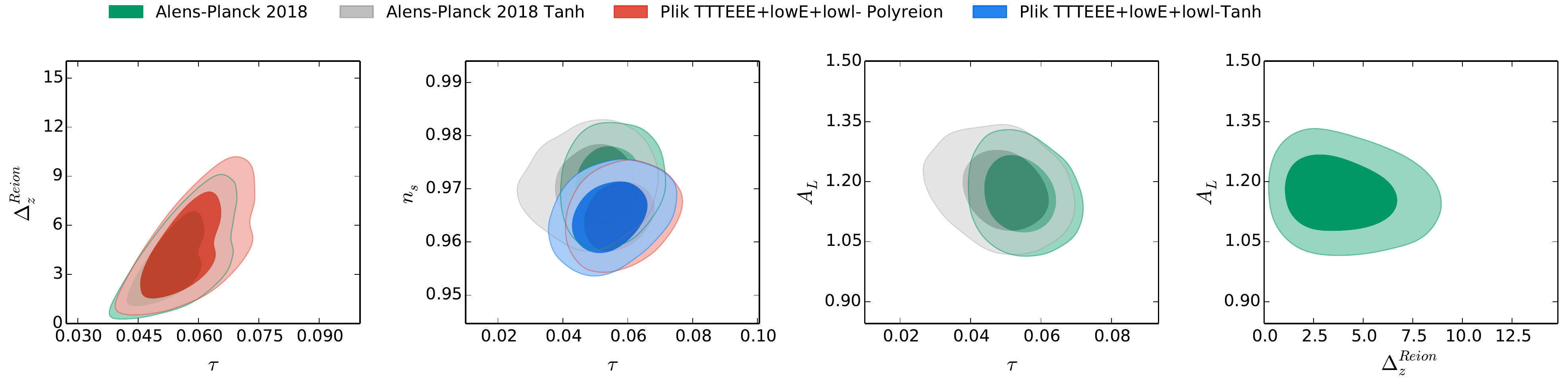}
\caption{\footnotesize\label{2DAl}Marginalized joint 68\% and 95\% CL regions for $(\tau,\Delta_z^{Reion})$, $(\tau,n_s)$,$(\tau,A_{L})$ and $(A_{L},\Delta_z^{Reion})$ compared with the hyperbolic tangent case and the LCDM.}
\end{figure}
When adding the $A_{L}$ parameter to the Poly-reion framework we obtain a slightly lower value with respect to the hyperbolic tangent, namely $A_{L}=1.17_{-0.07}^{+0.06}$ at 68\% CL, with similar uncertainty.
In Fig.~\ref{2DAl} we can see the drag of the Poly-reion model on the $A_{L}$ parameter by comparing the green contour with the analogous in hyperbolic tangent in red. The Poly-reion preference for higher optical depths together with the freedom in the duration of reionization, not allowed in the hyperbolic tangent (where the $\Delta_z^{Reion}$ is fixed), decreases the widening of the optical depth posterior and its pull towards lower values given by the inclusion of the lensing amplitude. In fact, with Poly-reion we have $\tau=0.054_{-0.008}^{+0.006}$ at 68\% CL whereas the hyperbolic tangent provides $\tau=0.049_{-0.007}^{+0.009}$ at 68\% CL. The uncertainties are almost comparable but the asymmetry in the error bars have opposite directions.
In Fig.~\ref{2DAl} we compare the $\Lambda$CDM+$A_{L}$ case with the standard $\Lambda$CDM results, we note that also for the Poly-reion model we have a pull of $A_{L}$ towards lower values of the optical depth as in the hyperbolic tangent, but the structure of Poly-reion somehow limits this pull.

\subsection{$\Lambda$CDM+curvature}

Another model which is sensitive to possible effects in the region of the acoustic peaks is the non-flat geometry.
Also in this case the Planck results have consistently preferred negative values of the curvature parameter, being $\Omega_k=-0.044_{-0.015}^{+0.018}$~\cite{Aghanim:2018eyx} the Planck 2018 results, corresponding to preference for a closed universe model. The addition of the lensing brings the value in full agreement with a flat universe but since we want to investigate the correlation of the spatial curvature with the Poly-reion model of reionization also in this case we exclude the lensing likelihood from the analysis.
\begin{figure}[!htb]
\includegraphics[width=\columnwidth]{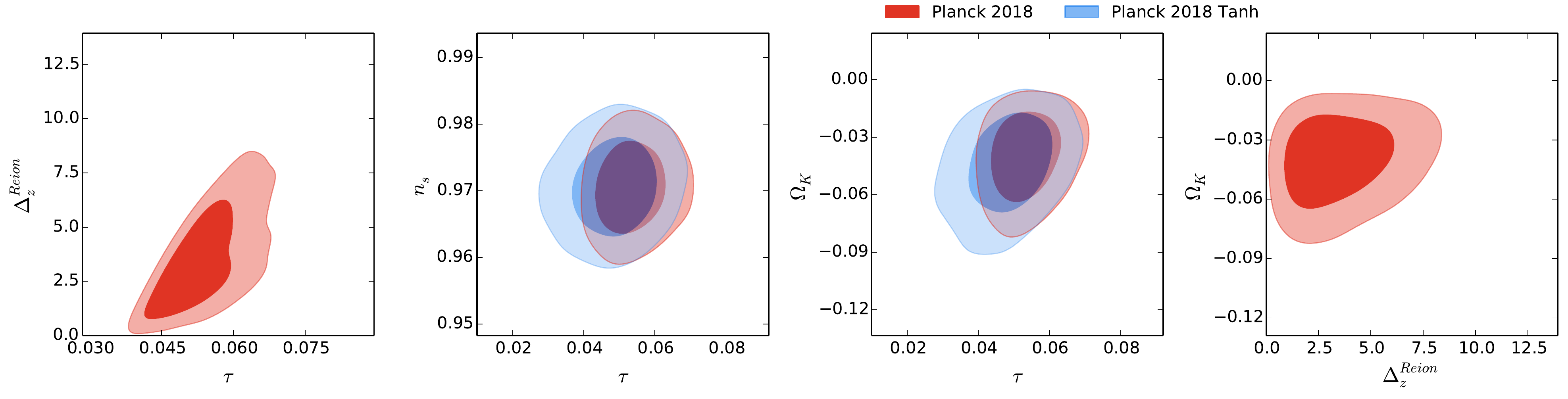}
\caption{\footnotesize\label{2DK}Marginalized joint 68\% and 95\% CL regions for $(\tau,\Delta_z^{Reion})$, $(\tau,n_s)$,$(\tau,\Omega_k)$ and $(\Omega_k,\Delta_z^{Reion})$ compared with the hyperbolic tangent case.}
\end{figure}
As expected the results show an analogous behaviour to the ones with the addition of $A_{L}$ as presented in Fig.~\ref{2DK}. We note how also for this case the pull towards lower values of the optical depth given by the addition of the curvature in the Poly-reion model $\tau=0.053_{-0.007}^{+0.006}$ is reduced with respect to the hyperbolic tangent case  $\tau=0.049_{-0.008}^{+0.007}$ with slightly smaller error bars for the Poly-reion case. The reduced pull towards lower values is reflected in a slightly higher value of the spatial curvature  $\Omega_K=-0.040_{-0.014}^{+0.017}$ at 68\% CL and slightly smaller error bars.

\subsection{Large scale power suppression}
\begin{figure}[!htb]
\includegraphics[width=\columnwidth]{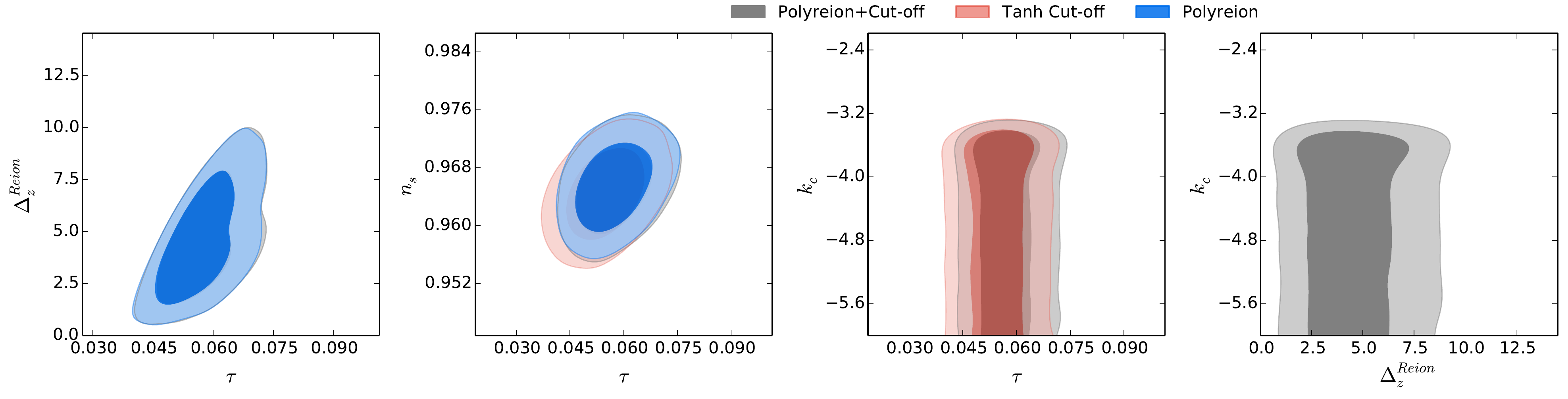}
\caption{\footnotesize\label{cutoff} Marginalized joint 68\% and 95\% CL regions for $(\tau,\Delta_z^{Reion})$, $(\tau,n_s)$,$(\tau,k_c)$ and $(k_c,\Delta_z^{Reion})$ in large scale power suppression model for the primordial power spectrum. The results for this model are compared with the standard Poly-reion case.}
\end{figure}

We now consider a possible variation to the initial power spectrum of primordial perturbations that wants to phenomenologically represent the power suppression observed on large angular scales in the CMB angular power spectra.
We model this effect by inserting a cut off into the primordial power spectrum for cosmological perturbations that takes the form~\cite{Akrami:2018odb}:
\begin{equation}
P(k)= P_0(k)\Bigg[1-\exp{-\Bigg(\frac{k}{k_c}\Bigg)^{\lambda_c}}\Bigg]\,,
\end{equation}
where $P_0(k)$ is the standard power law primordial power spectrum.
In order to reduce the parameter space, we consider the specific case of $\lambda_c=3.35$.
The results are presented in Fig.~\ref{cutoff}. We do not note any significant degeneracy among the Poly-reion parameters and the cut off pivot scale. 
The constraints on the Poly-reion parameters are the same as in the standard primordial power spectrum case showing again also for this extension that there are no strong degeneracies with Planck 2018 data.

\subsection{$\Lambda$CDM+DM annihilation}

We conclude our review of some of the main correlations of extended cosmological models with the Poly-reion by considering the contribution of dark matter annihilation to the cosmological environment~\cite{Aghanim:2018eyx,core:cosmoparam,Planck:2015Param,Kawasaki:2015peu,Lopez-Honorez:2013lcm,Galli:2013dna,Galli:2009zc,Chluba:2009uv}. 
Annihilating DM has a twofold effect on CMB anisotropies. On one side, the annihilated DM particles do not participate anymore to the matter budget of the Universe, decreasing the DM overall contribution to CMB and in particular to the acoustic peaks structure. On the other side, the annihilated DM has not disappeared but has been converted into additional radiation that is injected into the cosmological plasma. As for other forms of additional radiation or energy injections (e.g. the dissipation of primordial magnetic fields~\cite{Paoletti:2018uic}) this process modifies the ionization fraction evolution. This direct effect on the ionization fraction is the main reason why DM annihilation is degenerate with extended histories of reionization which phenomenologically reconstruct the ionization fraction evolution~\citep{HPFS18}.

The main free parameter describing the DM annihilation is the cross section usually parametrized as:  
\begin{equation}
p_{\mathrm ann}=\frac{f_{eff}\langle\sigma\nu\rangle}{m_\chi}
\end{equation}
where we have the particle mass at denominator and the thermally averaged
annihilation cross-section; $f_{eff}$ is the fraction of the energy
released and transferred to the plasma at redshift 600. 
We use the Cosmorec~\cite{cosmorec} extension which includes the injection of energy by dark matter annihilation in the plasma~\cite{Chluba:2009uv} that varies the parameter $\epsilon_0 f_d$ that is related to the $p_{\mathrm ann}$ by $p_{\mathrm ann}[\times 10^{-28} {\mathrm{cm^3/s/GeV}}]\sim 10 \epsilon_0 f_d (0.13/\omega_c h^2)^2$.
\begin{figure}[!htb]
\includegraphics[width=\columnwidth]{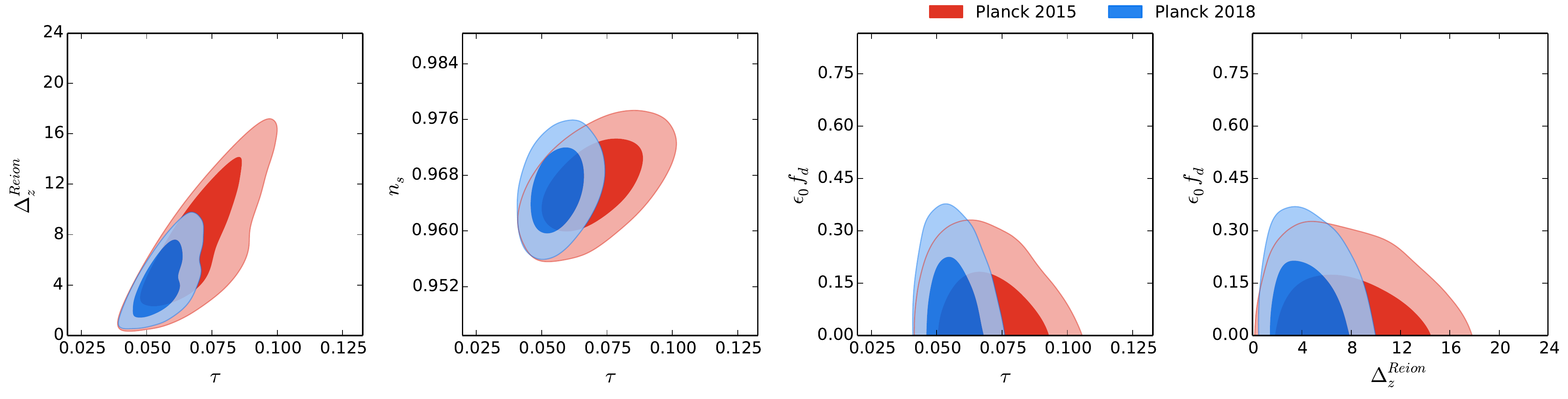}
\caption{\footnotesize\label{2DDMA}TMarginalized joint 68\% and 95\% CL regions for $(\tau,\Delta_z^{Reion})$, $(\tau,n_s)$,$(\tau,\epsilon_0 f_d)$ and $(\epsilon_0 f_d,\Delta_z^{Reion})$ compared with the hyperbolic tangent case.}
\end{figure}
In Fig.~\ref{2DDMA} we present the results, Planck 2018 data constrain $\epsilon_0 f_d<0.3$ that corresponds to  $p_{\mathrm ann}[\times 10^{-28} {\mathrm{cm^3/s/GeV}}]<4$ at 95\% CL The Poly-reion case is in agreement with, although slightly worse than, the hyperbolic tangent case which gives  $p_{\mathrm ann}[\times 10^{-28} {\mathrm{cm^3/s/GeV}}]<3.3$ at 95\% CL~\citep{Aghanim:2018eyx}.
If we compare the Poly-reion results with Planck 2018 and Planck 2015 shown in red in figure, we note that among all the Poly-reion extended cosmological models we have considered, the DM annihilation is the only one where we have a degradation of the constraint with respect to 2015 that gave $p_{\mathrm{ann}}[\times 10^{-28} {\mathrm{cm^3/s/GeV}}]<3$ at 95\% CL ~\citep{HPFS18}. Although the DM annihilation constraint is slightly degraded, the Planck 2018 determination of the optical depth, i.e. $\tau=0.056\pm 0.007$ at 68 \% CL, improves on the corresponding 2015 value in the DM annihilation model, i.e. $\tau=0.070_{-0.013}^{+0.012}$.
The degradation of the constraint on the DM annihilation cross section may be indeed driven by a combination of the lowering of the value of the optical depth  and its halved error bars that may have caused the correlation of the two parameters to move towards higher cross sections for DM.

\vspace{1cm}

In summary, the use of Planck 2018 baseline likelihoods remove most of the remaining degeneracies between the physics of reionization and the extended cosmological models we consider. In particular, the reionization parameters seem to decouple from the neutrino sector, the primordial tensor modes and a suppression of power at the largest scales and a small degeneracy seems to remain for the spatial curvature, a variable lensing amplitude and the contribution of annihilating dark matter. It is conceivable that the baseline Planck 2018 likelihood lowE, empirically obtained by simulations within $\Lambda$CDM, could contribute to the stability of the constraints on reionization parameters in extended cosmological models compared to $\Lambda$CDM. 

\section{Conclusions}

Thanks to the update of the large scale polarization from HFI 
and in particular the likelihood based on the 100x143 GHz cross angular power spectrum, 
Planck 2018 data made possible to investigate the history of reionization in great details. 
In the framework of the standard hyperbolic tangent model for the reionization history 
Planck 2018 data have led to a lower value of $\tau$ with 
a significant reduction in its uncertainty, i.e. $0.053\pm 0.009$ with only lowE and $0.0544\pm0.0073$ with the "baseline" combination including 
temperature, polarization and lensing~\cite{Aghanim:2018eyx}. In~\cite{Aghanim:2018eyx} 
are also presented the constraints on extended reionization by PCA \cite{PCAHu,PCA2} and 
Flexknot \cite{Millea:2018bko} approaches   
from the large angular scale E-mode polarization with the other cosmological parameters fixed.

We have studied the constraints with Planck 2018 data on the reionization history when 
it is modelled with Poly-reion~\citep{HS17} with flexible node~\cite{HPFS18}. This model 
by construction allows more freedom both in the duration of reionization and its phenomenology 
than the standard hyperbolic tangent model. 
The results on the $\Lambda$CDM model obtained by different combinations of Planck 2018 subsets 
of data show a great improvement with respect to the Planck 2015 results~\cite{HPFS18}. The Poly-reion
estimate of the optical depth is in agreement with the hyperbolic tangent 
model with a similar uncertainty, i.e. $\tau=0.0572^{+0.0064}_{-0.0075}$ at 68 \% CL. 
By using Poly-reion 
we find a significant reduction of the allowed duration of reionization compared to Planck 2015 data, 
i.e. $\Delta_z^{Reion}=4.6_{-2.5}^{+1.7}$ at 68\% CL, and that very early onsets 
are now disfavoured by Planck 2018 data.

We find consistent results within the Planck 2018 baseline likelihoods and between Planck 2018 baseline 
and alternative likelihoods, such as the LFI low-resolution joint $(T,Q,U)$ likelihood. 
We have also analyzed the update of the HFI
data processing from the SROLL2 map making algorithm~\cite{Delouis:2019bub}: we find 
$\tau=0.0600^{+0.0052}_{-0.0064}$, still consistent with the corresponding hyperbolic tangent 
result $\tau=0.0590^{+0.0058}_{-0.0068}$~\cite{Pagano:2019tci} 
(with a very small improvement due to the addition of the lensing likelihood in the Poly-reion results).

Models of the reionization history which allow for more freedom than the standard hyperbolic tangent may 
worsen the degeneracies between the reionization and cosmological model beyond $\Lambda$CDM. 
We have considered some of the main extensions to the $\Lambda$CDM 
that are known to be degenerate with the physics of reionization. 
We updated our previous results with Planck 2015~\cite{HPFS18} and we 
have also considered models with both neutrino number 
and mass allowed to vary, the addition of the CMB lensing amplitude $A_L$, a primordial tensor 
contribution and finally the spatial curvature. The results show that all the degeneracies are broken 
for both the neutrino physics sector and the primordial tensor contributions, providing constraints 
in perfect agreement with the one given by the hyperbolic tangent model. Some residual degeneracy 
is present instead for the lensing amplitude and the spatial curvature. In these models the use of 
the Poly-reion, for its tendency to increase the value of the optical depth, decreases the pull of 
the extended model towards lower values of the optical depth causing the degeneracy to shift the 
values of both the lensing amplitude and the spatial curvature towards $\Lambda$CDM case.
The DM annihilation is the only model where we have degradation of the constraint with Planck 
2018 with respect to Planck 2015 for the additional parameter of the model, we note however 
the improvement on the optical depth determination within this model. 

\begin{figure}[!htb]
\includegraphics[width=\columnwidth]{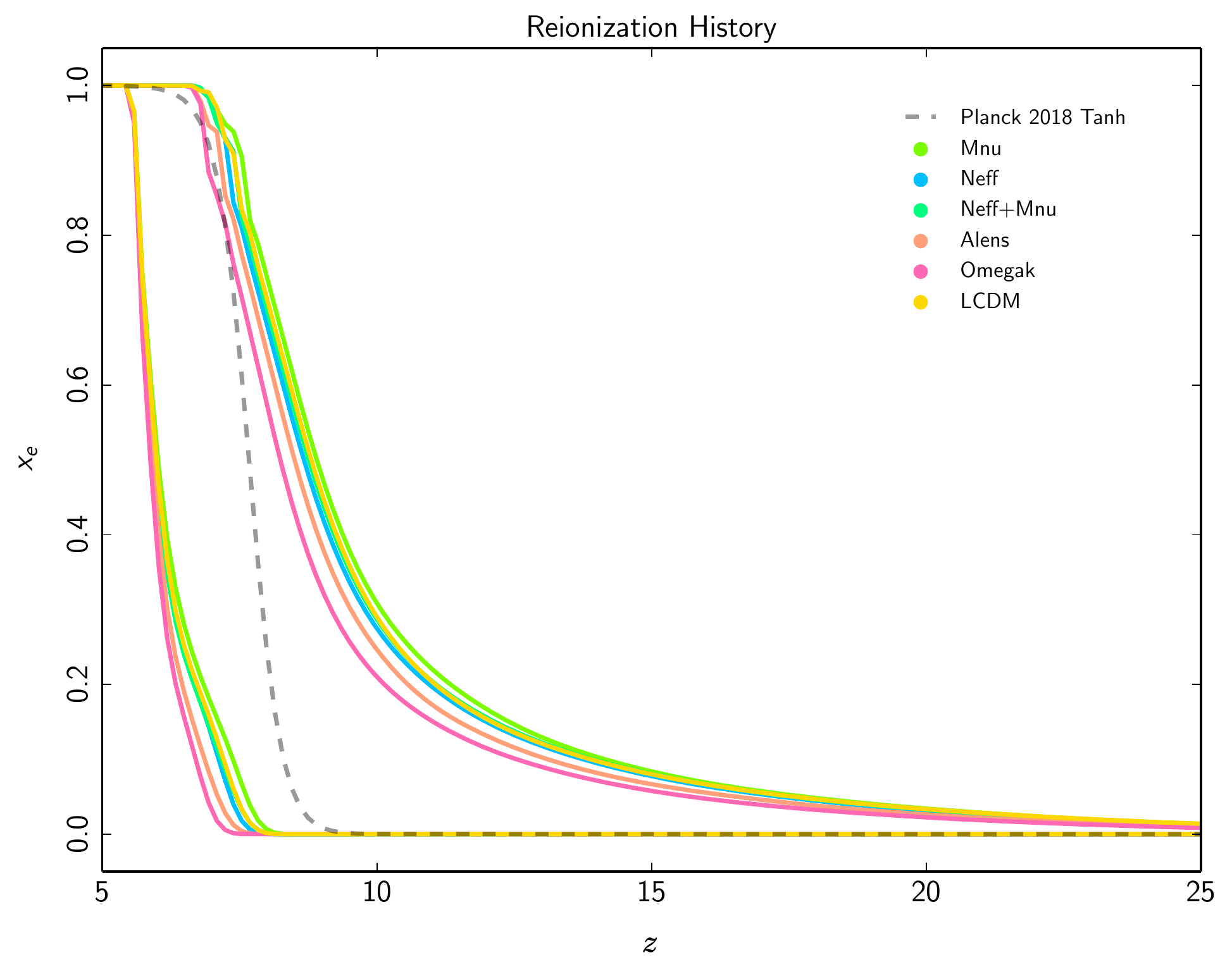}
\caption{\footnotesize\label{xeext}95\% CL bands for the ionization fraction as a function of 
redshift as reconstructed for a subset of the extensions to the $\Lambda$CDM considered in 
this work in different colours.}
\end{figure}

In Figure~\ref{xeext} we summarize the constraints on the Poly-reion reionization history 
in the framework of the cosmological models we have considered by 
showing the evolution of the ionization fractions. The comparison with the Poly-reion $\Lambda$CDM 
case reflects the results on the cosmological parameters with the neutrino physics being in perfect 
agreement whereas the spatial curvature and lensing amplitude prefers slightly more recent onsets of 
the reionization in agreement with a lower value of the optical preferred by these models.

The understanding of the history of reionization is one of the main scientific targets 
of future CMB experiments from space 
thanks to the cosmic variance limited measurement of the E-mode polarization on a wide range of multipoles, 
including the reionization bump at the lowest multipoles.
Future experiments like the LiteBIRD satellite~\cite{Hazumi:2019lys} 
will provide a cosmic variance limited measurement of the optical depth, 
$\sigma (\tau) \sim 0.002$ and provide tight constraints on the duration 
of reionization see~\cite{HPFS18} for some detailed forecasts.

\acknowledgments
We  would like to thank Luca Pagano for the useful discussions and for helping in the use of SROLL2 likelihood.
DP and FF acknowledge financial support by ASI Grant 2016-24-H.0. This research used computational resources provided by INAF OAS Bologna and by CINECA under the agreement with INFN.
DKH has received fundings from the European Union’s Horizon 2020 research and innovation programme under the Marie Sklodowska-Curie grant agreement  No. 664931.

\end{document}